\documentclass[12pt]{iopart}

\usepackage{iopams}
\usepackage{graphicx}
\usepackage{subfigure}
\usepackage{color}
\usepackage{hyperref}
\usepackage{cite}
\usepackage{mcite}

\begin{document}

\title{Fate of superconductivity in disordered Dirac and semi-Dirac semimetals}

\author{Jing-Rong Wang$^{1,2}$, Guo-Zhu Liu$^{2,4}$, and
Chang-Jin Zhang$^{1,3,4}$ \footnote[0]{$^4$ Author to whom any
correspondence should be addressed.}}
\address{$^1$Anhui Province Key Laboratory of Condensed Matter
Physics at Extreme Conditions, High Magnetic Field Laboratory of the
Chinese Academy of Sciences, Hefei, Anhui 230031, China}
\address{$^2$Department of Modern Physics, University of Science and
Technology of China, Hefei, Anhui 230026, China}
\address{$^3$Institute of Physical Science and Information
Technology, Anhui University, Hefei, Anhui 230601, China}

\eads{\mailto{gzliu@ustc.edu.cn}; \mailto{zhangcj@hmfl.ac.cn}}

\begin{abstract}
The influence of weak disorder on the superconductivity in ordinary
metals can be formally described by the Abrikosov-Gorkov
diagrammatic approach. The vertex correction is ignored in this
approach because an inequality $k_F l \gg 1$, where $k_F$ is the
Fermi momentum and $l$ mean free path, is satisfied in ordinary
metals with a large Fermi surface. In a Dirac semimetal that has
discrete Fermi points, this inequality may break down even for
arbitrarily weak disorder since $k_F \rightarrow 0$, and thus the
vertex correction could be important. We incorporate the vertex
correction into the self-consistent equations of the superconducting
gap and the disorder scattering rate, and then apply the generalized
approach to study how $s$-wave superconductivity is affected by
random chemical potential in two- and three-dimensional Dirac
semimetals, as well as two-dimensional semi-Dirac semimetal. In the
clean limit, superconductivity is formed only when the pairing
interaction strength is greater than some critical value in these
materials. Adding random chemical potential to the system promotes
superconductivity by generating a finite fermionic density of states
at the Fermi level. In three-dimensional Dirac semimetal, the
critical attraction strength is reduced by weak disorder, but
remains finite. In the other two cases, superconductivity is induced
by arbitrarily weak attraction. Including the vertex correction does
not change these qualitative results, and actually could further
promote superconductivity in the weak-attraction regime. Bilayer
graphene is quite special in that its zero-energy density of states
is nonzero despite the existence of Fermi points. Due to this
peculiar property, superconductivity is always slightly suppressed
by random chemical potential, and the impact of vertex correction is
nearly negligible.
\end{abstract}




\maketitle


\section{Introduction}

The advance made in the fabrication of various types of semimetals
(SMs) has stimulated a huge number of research works in the past
decade \cite{CastoNeto09, Peres10, Sarma11, Kotov12, Hasan10, Qi11,
Hasan11, Vafek14, Wehling14, Armitage18, Burkov16, Yan17, Hasan17,
Burkov18, Weng16, FangChen16}. Graphene, made of one layer of carbon
atoms on a honeycomb lattice, is a typical two-dimensional (2D)
Dirac SM (DSM) and has been extensively studied \cite{CastoNeto09,
Peres10, Sarma11, Kotov12}. Moreover, the gapless surface electronic
state of three-dimensional (3D) topological insulators can also be
regarded as 2D DSM \cite{Hasan10, Qi11, Hasan11}. Recently, a stable
3D DSM state that is protected by crystal symmetry was found in
Na$_{3}$Bi and Cd$_{3}$As$_{2}$ \cite{Vafek14, Wehling14,
Armitage18}. In addition, Weyl SM (WSM), in which the low-energy
fermionic excitations display linear dispersion around pairs of
nodes with opposite chirality, was observed in TaAs, NbAs, TaP, and
NbP by the angle-resolved photoemission spectroscopy (ARPES)
\cite{Armitage18, Burkov16, Yan17, Hasan17, Burkov18}. There also
exist other types of SMs, such as 3D nodal line SM (NLSM)
\cite{Weng16, FangChen16, Sur16, Roy17NLSM}, 2D semi-DSM
\cite{Hasegawa06, Dietl08, Goerbig08, Montambaux09B, Kobayashi11,
Dolui15, Yuan16, Baik15, Wunsch08, Lim12, Tarruell12, Bellec13,
Kim15, Pardo09, Banerjee09, Banerjee12, Kamal15, CanWang16, Dora13,
Pyatkovskiy16, Isobe16, Cho16, WangLiuZhang17, AhnYang17, Uchoa17,
RoyFoster18, Carpentier13, ZhaoPengLu16}, 3D double-WSM
\cite{XuFang11, Fang12, Yang14A, Huang16, Lai15, Jian15,
WangLiuZhang16, BeraRoy16, Sbierski17PRB, Roy17, Ahn16, Ahn17,
Park17}, 3D triple-WSM \cite{Yang14A, WangLiuZhang16, Roy17, Ahn16,
Ahn17, Park17, LiuZunger17, Zhang16, WangLiuZhang17B}, 3D
anisotropic-WSM \cite{Yang14A, Yang13, Yang14B, Moon14}, and 3D
Luttinger SM \cite{Herbut14, Janssen15, JanssenSeries, Boettcher16,
Boettcher17}.

Most of these SMs share a common feature: the Fermi surface is
composed of a number of zero-dimensional points. This implies that
the fermion density of states (DOS) vanishes at the Fermi level,
which is different from conventional metals whose DOS takes a finite
value at the Fermi level. This difference is responsible for many
intriguing properties of SM materials that cannot occur in ordinary
metals.

Cooper pairing instability in 2D DSM has attracted considerable
interest in recent years \cite{CastoNeto01, Uchoa05, Zhao05,
Marino06, Uchoa07, BlackSchaffer07, Kopnin08, Honerkamp08,
Gonzalez08, Loktev09, Roy10, Roy13, Santos10, Ito11, Ito12,
Nandkishore12, Nandkishore13, Potirniche14, Dietel14, Li14, She14,
WangJing17}. It has been proposed that phonon or plasmon may mediate
an effective attraction between Dirac fermions \cite{Kotov12,
Uchoa05, Uchoa07}. For an undoped 2D DSM, superconductivity can be
realized only when the net attraction is sufficiently strong
\cite{CastoNeto01, Uchoa05, Zhao05, Marino06, Uchoa07,
BlackSchaffer07, Kopnin08, Honerkamp08, Gonzalez08, Loktev09, Roy10,
Roy13, Santos10, Ito11, Ito12, Nandkishore12, Nandkishore13,
Potirniche14, Dietel14, Li14, She14, WangJing17}. This is quite
different from the conventional metal superconductors in which even
an arbitrarily weak attraction suffices to trigger Cooper pairing
\cite{GiulianiBook, ColemanBook, Shankar94}. The difference is owing
to the fact that the fermion DOS vanishes at zero energy, namely
$\rho(0) = 0$. Cooper pairing instability may be achieved in other
SMs, including 2D semi-DSM \cite{Uchoa17, RoyFoster18}, 3D DSM
\cite{Roy16}, 3D WSM \cite{Meng12, Maciejko14}, 3D NLSM \cite{Sur16,
Roy17NLSM}, and 3D Luttinger SM \cite{Janssen15, Boettcher16,
Boettcher17}. In these materials, there is also a threshold value
for the strength of attraction.

In terms of industrial and commercial applications, the existence of
a threshold for net attraction appears to be a negative result
because it makes it difficult to realize intrinsic superconductivity
in undoped DSMs. However, from the perspective of theoretical study,
it provides us with a unique opportunity to investigate various
novel properties that do not exist in ordinary metal
superconductors. The critical value of attraction defines a genuine
quantum critical point (QCP) between the SM and superconducting (SC)
phases at zero temperature, and the system exhibits a wealth of
attractive quantum critical behaviors around such QCP
\cite{Boettcher16, Ponte14, Grover14, Witczak-Krempa16, Zerf16,
Jian15SUSY, Jian17}. For instance, an effective space-time
supersymmetry was recently argued to emerge at this QCP at low
energies \cite{Ponte14, Grover14, Witczak-Krempa16, Zerf16,
Jian15SUSY, Jian17}.

An interesting problem is how superconductivity formed in various
DSMs is affected by weak disorder. Among the possible disorders,
random chemical potential is most frequently encountered in
realistic materials \cite{Evers08, Sarma11, Syzranov18Review}. In
the case of conventional metal superconductors, Anderson theorem
states that weak random chemical potential does not alter the SC gap
$\Delta$ and the critical temperature $T_c$ if the pairing is
$s$-wave \cite{Anderson59, Gorkov08, Belitz94, Balatsky06, Lee85}.
This result can be obtained by the diagrammatic approach developed
by Abrikosov and Gorkov (AG) \cite{Gorkov08, AbrikosovGorkov59}. For
such result to be valid, an important precondition is that the
low-energy fermion DOS is not sensitive to weak disorder
\cite{Balatsky06, Gorkov08}. While this condition is satisfied in
ordinary metals with a large Fermi surface, it breaks down in DSMs
that have only isolated Fermi points. Renormalization group (RG)
analysis showed that random chemical potential is a relevant
perturbation to 2D DSM \cite{Nandkishore13, WangJing17, Ludwig94,
Ostrovsky06, Evers08, Foster12}, thus arbitrarily weak disorder
eventually flows to the strong coupling regime and then drives the
system to enter into a compressible diffusive metal (CDM) state, in
which disorder generates a finite scattering rate $\Gamma$. In
addition, the fermions acquire a finite zero-energy DOS $\rho(0)$
that is a function of $\Gamma$ \cite{Nandkishore13, WangJing17,
Ludwig94, Evers08, Foster12, Fradkin86, Durst00, Ostrovsky06,
Katanin13}. It is obvious that the zero-energy DOS of the 2D DSM is
very sensitive to weak disorder, in stark contrast to the case of
ordinary metals. As a result, weak random chemical potential might
have significant influence on $\Delta$ and $T_c$. The question is
whether superconductivity is enhanced or suppressed.

In a recent work \cite{WangJing17}, the influence of random gauge
potential and random mass on superconductivity in 2D DSM was studied
by using the perturbative RG method. It was found there that the
critical attraction for Cooper pairing is increased by random gauge
potential or random mass \cite{WangJing17}. The strength parameter
of random gauge potential can be fixed at a small value, whereas the
strength parameter of random mass flows to zero logarithmically as
the zero-energy limit is reached \cite{Ludwig94, Ostrovsky06,
Evers08, Foster12, WangJing17}. Therefore, the perturbative RG
analysis is reliable in these two cases.

If random chemical potential and pairing interaction are considered
simultaneously, RG analysis revealed \cite{WangJing17} that both of
the two strength parameters tend to diverge at some finite energy
scale. This indicates that the system becomes unstable and will be
turned into a new phase. In such a strong coupling regime, the RG
method cannot be used to determine whether the system enters into a
CDM phase or a SC phase. To address this issue, we need to analyze
the possible generation of disorder scattering rate $\Gamma$ and the
possible generation of SC gap $\Delta$. These two processes strongly
affect each other. Therefore, the quantities $\Gamma$ and $\Delta$
should be calculated in a self-consistent and unbiased way.

Dyson-Schwinger equations (DSEs) provide a suitable framework to
treat disorder scattering and Cooper pairing on an equal footing.
The essence of this approach is to solve the self-consistently
coupled DSEs for $\Gamma$ and $\Delta$. In the most generic form,
DSEs are complete and contain all the information about disorder
scattering and Cooper pairing. In practice, however, it is always
necessary to properly truncate the complete set of DSEs. The
original AG method and its generalization to be presented below in
this work are actually two different truncations of the complete set
of DSEs.

Nandkishore \emph{et al.} \cite{Nandkishore13} investigated the
effect of random chemical potential on $s$-wave superconductivity by
taking the surface state of 3D topological insulator as an example.
After solving the mean-field gap equation, which is a simplified
version of AG method, in combination with a RG analysis
\cite{Nandkishore13}, they found that superconductivity can be
induced by arbitrarily weak attraction. Subsequently, Potirniche
\emph{et al.} \cite{Potirniche14} analyzed the interplay of
superconductivity and random chemical potential by considering a
Hubbard model defined on honeycomb lattices with an on-site
attractive coupling $U$ by means of self-consistent Bogoliubov-de
Gennes  equation method \cite{Potirniche14}. In the clean limit,
they argued that the system remains a SM if $U$ is smaller than
certain critical value $U_{c}$, but becomes SC when $U$ exceeds
$U_{c}$. Adding disorder to the clean system results in a
complicated behavior of superconductivity. In the strong coupling
regime $U > U_{c}$, disorder can suppress superconductivity. In the
weak coupling regime $U < U_{c}$, weak disorder enhances
superconductivity, but strong disorder eventually destroys
superconductivity.

In Ref.~\cite{WangJing17}, the diagrammatic AG method was applied to
study the impact of random chemical potential on superconductivity
in the context of a 2D DSM, yielding results that are qualitatively
consistent with those of Nandkishore \emph {et al.}
\cite{Nandkishore13} and Potirniche \emph{et al.}
\cite{Potirniche14}. However, here we emphasize that the
applicability of the original AG method to 2D DSM is actually
questionable \cite{Nandkishore13, WangJing17}, because it entirely
neglects the vertex correction to the fermion-disorder coupling.
Such correction would be small if the inequality $k_F l \gg 1$,
where $k_F$ is the Fermi momentum and $l$ the mean free path, is
satisfied. Since the disorder scattering rate $\Gamma$ is inversely
proportional to $l$, the above inequality can be re-expressed as
$k_F \gg \Gamma$. In ordinary metals with a finite Fermi surface,
the Fermi momentum is usually large and the scattering rate caused
by weak disorder is small, thus this inequality is generically
satisfied. However, 2D DSM contains only discrete Fermi points,
which means that $k_F \rightarrow 0$. Additionally, random chemical
potential is a relevant perturbation in 2D DSM and always induces a
finite disorder scattering rate $\Gamma$. In this case the
inequality $k_{F} \gg \Gamma$ breaks down, so the vertex correction
may be very important. It is interesting to verify whether random
chemical potential still promotes superconductivity after including
the vertex correction. To obtain a reliable relation between
superconductivity and random chemical potential, one should go
beyond the original AG approximation and study the role of vertex
correction. This is the first motivation of this work.

We notice an important principle that the impact of random chemical
potential in various SMs is closely related to the dimensionality
and the specific fermion dispersion. For example, weak random
chemical potential is irrelevant in a 3D DSM/WSM, but becomes
relevant if its strength is sufficiently large. Accordingly, for a
3D DSM, there is a QCP between the SM and CDM phases with varying
disorder strength \cite{Goswami11, Roy14, Syzranov18Review, Fu17,
Sbierski17, Ominato15, Sinner17}. This behavior is apparently
different from the case of 2D DSM, where an arbitrarily weak random
chemical potential leads to the CDM transition. For 2D semi-DSM
\cite{Carpentier13, ZhaoPengLu16}, in which the fermion dispersion
is linear along one direction and quadratic along the other
direction, the CDM state is achieved by an arbitrarily weak
disorder, analogous to 2D DSM. The second motivation of this work is
to examine whether the conclusion obtained previously in the case of
2D DSM is applicable to other SMs.

We will show that arbitrarily weak attraction suffices to induce an
$s$-wave superconductivity in 2D DSM when random chemical potential
is present. If the attraction is weak, the magnitude of SC gap
increases with growing disorder strength, but begins to decrease
when the disorder strength becomes large enough. For relatively
strong attraction, the SC gap is always suppressed by the increasing
disorder strength. We include the vertex correction into the
self-consistent equations for the SC gap and the disorder scattering
rate, and show that the above behavior is not altered qualitatively
by the vertex correction. In the case of weak attraction and weak
disorder, the magnitude of SC gap is further amplified once the
vertex correction is incorporated.

The generalized AG method can be also applied to other SM systems to
examine whether or not the vertex correction plays a vital role. To
make our analysis more comprehensive, and also to examine how
superconductivity relies on the dimensionality and the energy
dispersion of the fermion excitations, we then study the fate of
superconductivity in disordered 3D DSM and 2D semi-DSM, which both
have a point-touching band structure.

In the case of slightly disordered 3D DSM, the critical pairing
interaction strength $g_c$ is smaller than that obtained in the
clean limit, but remains finite. Thus, there is still a SC QCP, at
which interesting quantum critical phenomena might occur. When the
3D DSM contains strong random chemical potential, the critical value
$g_c$ vanishes, and superconductivity is formed by arbitrarily weak
pairing interaction. An apparent conclusion is that
superconductivity is promoted by random chemical potential in 3D
DSM. In a 2D semi-DSM, the disorder effect on superconductivity is
very similar to 2D DSM: superconductivity occurs only when the
pairing interaction strength exceeds a threshold $g_c$ in the clean
limit, but is triggered by arbitrarily weak attraction when random
chemical potential is introduced. In both of these two systems,
including vertex correction results in more significant enhancement
of SC gap in the weak-attraction regime.

Comparing to the above three SMs, the bilayer graphene is special in
that its Fermi surface is composed of discrete points but the DOS is
finite at the Fermi level. We find that $s$-wave superconductivity
in bilayer graphene is always slightly suppressed by random chemical
potential, and that the vertex correction plays a minor role. This
result is qualitatively different from the other three kinds of SM.
The difference stems from the fact that the zero-energy DOS is
nonzero only in bilayer graphene but vanishes otherwise.

The rest of paper is organized as follows. In
section~\ref{Sec:Hamiltonian}, we present the model Hamiltonian for 2D
DSM, analyze the influence of disorder on superconductivity by means
of the AG method, and examine the role played by the vertex
correction. The disorder effect on superconductivity in 3D DSM and
2D semi-DSM is studied in section~\ref{Sec:3DDSM} and
section~\ref{Sec:2DSemiDSM}, respectively. In
section~\ref{Sec:BilayerGraphene}, we apply the same approach to the
case of bilayer graphene. In section~\ref{Sec:Discussions}, we give a
remark on several related questions, including truncation of DSEs, the effect of rare region, and Anderson localization. We summarize
our main results in section~\ref{Sec:SummaryAndDiscussion}.

\section{Superconductivity in 2D Dirac semimetal\label{Sec:Hamiltonian}}

In this section, we consider the case of 2D DSM and study the impact
of random chemical potential on superconductivity by means of the AG
approach and its proper generalization. The framework employed in
this section is quite general and will be utilized in the following
several sections to study the fate of superconductivity in 3D DSM,
2D semi-DSM, and bilayer graphene.

\subsection{Model Hamiltonian}

We consider a single species of massless Dirac fermions that emerge,
for instance, on the surface of a 3D topological insulator.
Following the notations adopted in Ref.~\cite{Nandkishore13}, we
write the Hamiltonian in the form
\begin{eqnarray}
\fl H &=& \sum_{\mathbf{k}}\psi_{\mathbf{k}}^{\dag}\left(-\mu\sigma_{0}
+ v k_{x}\sigma_{1} + v k_{y}\sigma_{2}\right) \psi_{\mathbf{k}}
-g\sum_{\mathbf{k},\mathbf{q}} \psi_{\mathbf{k}}^{\dag}
\left(-i\sigma_{2}\right) \psi_{-\mathbf{k}}^{\dag}
\psi_{\mathbf{q}}i\sigma_{2} \psi_{-\mathbf{q}},
\end{eqnarray}
where $\mu$ is the chemical potential and $g$ is the coupling
constant for the BCS-type attractive interaction. At a finite $\mu$,
the DSM has a finite Fermi surface, and many of its
low-energy properties are very similar to ordinary metals. In the
following, we only consider the most interesting case of zero
chemical potential, $\mu = 0$, corresponding to undoped SM.
The spinor field is defined as $\psi_{\mathbf{k}} =
(\begin{array}{cc} c_{\uparrow \mathbf{k}}, & c_{\downarrow
\mathbf{k}} \end{array})^{T}$, whose conjugate is
$\psi_{\mathbf{k}}^{\dag} = \left(\begin{array}{cc}
c_{\uparrow\mathbf{k}}^{\dag}, & c_{\downarrow\mathbf{k}}^{\dag}
\end{array} \right)$. Moreover, $\sigma_{1,2}$ are standard Pauli
matrices, and $\sigma_{0}$ is identity matrix. The SC order
parameter is defined as
\begin{eqnarray}
\Delta = g\sum_{\mathbf{k}} \langle \psi_{\mathbf{k}}(i\sigma_{2})
\psi_{-\mathbf{k}}\rangle.
\end{eqnarray}
The system enters into a SC phase when $\Delta$ acquires a nonzero
value.

Our analysis starts from the partition function
\begin{eqnarray}
Z = \int D\left[\psi^{\dag},\psi\right]
\exp\left(-\int_{0}^{\beta}d\tau\int d^2\mathbf{x}
L\left[\psi^{\dag},\psi\right]\right),
\end{eqnarray}
where $\beta = 1/T$ and the Lagrangian is related to the Hamiltonian
$H$ via the Legendre transformation
\begin{eqnarray}
L = \int\frac{d^2\mathbf{k}}{(2\pi)^{2}}
\psi_{\mathbf{k}}^{\dag}\partial_{\tau}\psi_{\mathbf{k}} + H.
\end{eqnarray}
To express the action in a more compact form, it is convenient to
introduce a four-component Nambu spinor:
\begin{eqnarray}
\Psi = \left(
\begin{array}{cc}
\psi_{\mathbf{k}},\psi_{-\mathbf{k}}^{\dag}
\end{array}
\right)^{T}.
\end{eqnarray}
After decoupling the quartic attractive interaction by means of
Hubbard-Stratonovich transformation, we can re-write the partition
function as
\begin{eqnarray}
Z &=& \int D\left[\Psi^{\dag},\Psi,\Delta^{*},\Delta\right]
\exp\left(-\int_{0}^{\beta}d\tau\int d^2\mathbf{x}
L\left[\Psi^{\dag},\Psi,\Delta^{*},\Delta\right]\right),
\end{eqnarray}
where $L$ takes the form
\begin{eqnarray}
L = T\sum_{\omega_{n}} \sum_{\mathbf{k}}
\Psi_{\omega_{n},\mathbf{k}}G_{\omega_{n},\mathbf{k}}^{-1}
\Psi_{\omega_{n},\mathbf{k}} - \frac{\left|\Delta\right|^{2}}{g}.
\end{eqnarray}
In the above expression, we have defined a fermion propagator
$G_{\omega_{n},\mathbf{k}} \equiv G(\omega_{n},\mathbf{k})$ that is
given by
\begin{eqnarray}
G_{\omega_{n},\mathbf{k}}^{-1} = \left(
\begin{array}{cccc}
i\omega_{n} & vk_{+} & 0 & \Delta
\\
vk_{-} & i\omega_{n} & -\Delta & 0
\\
0, &-\Delta^{*} & i\omega_{n} & vk_{-}
\\
\Delta^{*} & 0 & vk_{+} & i\omega_{n}
\end{array}
\right),
\end{eqnarray}
where $k_{\pm}=k_{x}\pm ik_{y}$. Within the Matsubara formalism, the
fermion frequency is $\omega_{n} = (2n+1)\pi T$ with $n$ being
integers. The above expression of partition function is consistent
with Ref.~\cite{Nandkishore13}.

\begin{figure}
\center
\includegraphics[width=3.8in]{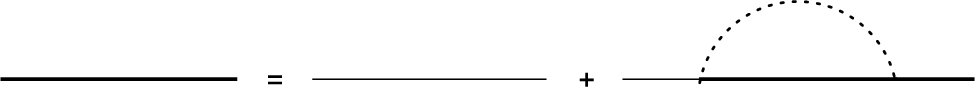}
\caption{Feynman diagram for the fermion self-energy. Thin line
represents the free fermion propagator, and thick line the full
fermion propagator. Dotted line represents disorder scattering. The
vertex correction is neglected in the original AG method, which is
valid if $k_{F}l \gg 1$. \label{Fig:FeynmanAG}}
\end{figure}

\subsection{Clean case\label{Sec:GapCleanCase}}

The SC gap equation can be derived by integrating over
fermion fields $\Psi$ and $\Psi^{\dag}$. For this purpose, the
partition functions is further written as
\begin{eqnarray}
Z &=& \int\mathcal{D}\Psi^{\dag}\mathcal{D}\Psi e^{S}
= \prod_{\omega_{n}}\prod_{\mathbf{k}}\int\mathcal{D}
\Psi^{\dag}\mathcal{D}\Psi e^{ \Psi_{\omega_{n},\mathbf{k}}^{\dag}
\beta G_{\omega_{n},\mathbf{k}}^{-1}\Psi_{\omega_{n},\mathbf{k}}}
e^{-\int_{\tau,\mathbf{x}}\frac{\left|\Delta\right|^{2}}{g}},
\end{eqnarray}
where $\int_{\tau,\mathbf{x}}\equiv\int_{0}^{\beta} d\tau \int
d^{2}\mathbf{x}$. Performing the functional integration over $\Psi$
and $\Psi^{\dag}$ yields
\begin{eqnarray}
Z = \prod_{\omega_{n}}\prod_{\mathbf{k}}\beta^{4}\det
\left(G_{\omega_{n},\mathbf{k}}^{-1}\right) e^{-\int_{\tau,\mathbf{x}}
\frac{\left|\Delta\right|^{2}}{g}}.
\end{eqnarray}
It is then easy to get
\begin{eqnarray}
\ln Z = \sum_{\omega_{n}}\sum_{\mathbf{k}} \ln\left[\beta^{4}\det
\left(G_{\omega_{n},\mathbf{k}}^{-1}\right)\right] -
\int_{\tau,\mathbf{x}}\frac{\left|\Delta\right|^{2}}{g}.
\end{eqnarray}
Through direct calculation, we obtain
\begin{eqnarray}
\det G_{\omega_{n},\mathbf{k}}^{-1} =
\left(\omega_{n}^2 + v^{2}k^2
+ \left|\Delta\right|^2\right)^{2}.
\end{eqnarray}

For a sample of volume $V$, the free energy density is
\begin{eqnarray}
\fl f &=& \frac{F}{V}=-\frac{1}{\beta V}\ln Z = -\frac{1}{\beta
V}\sum_{\omega_{n}}\sum_{\mathbf{k}} \ln\left[\beta^{4}
\left(\omega_{n}^2 + v^{2}k^2 + \left|\Delta\right|^2
\right)^{2}\right] + \frac{\left|\Delta\right|^{2}}{g}.
\end{eqnarray}
Making variation with respective to infinitesimal change of
$\Delta$, $\frac{\delta f}{\delta \Delta}=0$, we finally obtain the
gap equation:
\begin{eqnarray}
&& 2T\sum_{\omega_{n}} \int
\frac{d^2\mathbf{k}}{\left(2\pi\right)^2}
\frac{1}{\omega_{n}^2+v^{2}k^{2} + \Delta^2} = \frac{1}{g}.
\end{eqnarray}
We have already fixed the phase factor of the gap function $\Delta$,
and in the following will take $\Delta$ as a real variable. At zero
temperature $T = 0$, the gap equation becomes
\begin{eqnarray}
2\int\frac{d\omega}{2\pi}\frac{d^2\mathbf{k}}{(2\pi)^2}
\frac{1}{\omega^2 + v^{2}k^{2}+\Delta^2}=\frac{1}{g}.
\end{eqnarray}
Performing the integration of $\omega$ and $k$, we obtain
\begin{eqnarray}
\frac{g}{2\pi v^2} \left(\sqrt{v^2\Lambda^2 +
\Delta^2}-\Delta^2\right) = 1,
\end{eqnarray}
where $\Lambda$ is the cutoff of the momentum. Setting $\Delta = 0$
yields the critical coupling $g_{c0} = 2\pi v/\Lambda$. The gap
$\Delta$ acquires a nonzero value only when $g > g_{c0}$, and there
is a SM-SC QCP at $g = g_{c0}$ \cite{Uchoa05, Marino06, Kopnin08,
Nandkishore13, WangJing17}.

\begin{figure}
\center
\includegraphics[width=2.2in]{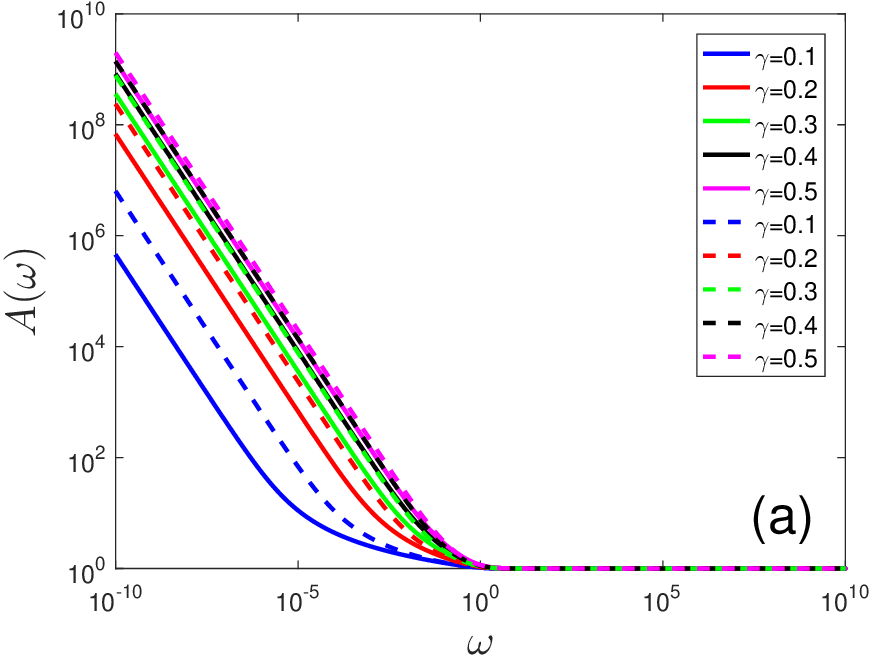}
\includegraphics[width=2.2in]{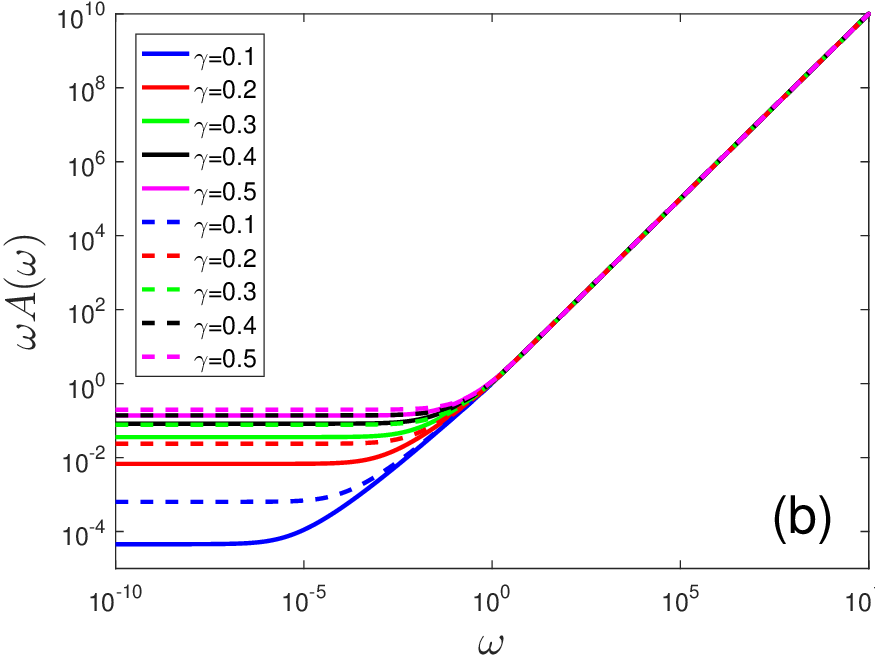}
\caption{(a) Dependence of $A$ and (b) $\omega A$ on $\omega$ at
different values of $\gamma$ for 2D DSM. Vertex correction is
neglected for solid lines, but incorporated for dashed lines. Here, we take $\Delta =
0$, corresponding to non-SC phase. \label{Fig:AFun}}
\end{figure}

\subsection{Analysis by AG method without vertex correction}
\label{Sec:DerivationAG}

In the presence of random chemical potential, the dynamics of Dirac
fermions is modified due to the disorder scattering. Consequently,
the fermion propagator can be expressed by the matrix
\begin{eqnarray}
\left(G_{\omega_{n},\mathbf{k}}^{F}\right)^{-1} = \left(
\begin{array}{cccc}
A_{1} i\omega_{n} & A_{2}vk_{+} & 0 & A_{3}\Delta
\\
A_{2}vk_{-} & A_{1}i\omega_{n} & -A_{3}\Delta & 0
\\
0, &-A_{3}\Delta^{*} & A_{1} i\omega_{n}  & A_{2}vk_{-}
\\
A_{3}\Delta^{*} & 0 & A_{2}vk_{+} & A_{1} i\omega_{n}
\end{array}
\right),
\end{eqnarray}
where $A_{i}\equiv A_{i}(\omega_{n})$ with $i=1,2,3$ are the
renormalization functions induced by the disorder scattering. The
propagators $G_{\omega_{n},\mathbf{k}}^{F}$ and
$G_{\omega_{n},\mathbf{k}}$ are connected with each other through
the DS
\begin{eqnarray}
\left(G_{\omega_{n},\mathbf{k}}^{F}\right)^{-1} =
G_{\omega_{n},\mathbf{k}}^{-1}-\Sigma_{\mathrm{dis}}(\omega_{n}),
\end{eqnarray}
where
\begin{eqnarray}
\Sigma_{\mathrm{dis}}(\omega_{n}) = n_{\mathrm{imp}}u^2\int
\frac{d^2\mathbf{k}}{(2\pi)^{2}} G_{\omega_{n},\mathbf{k}}^{F},
\end{eqnarray}
where $n_{\mathrm{imp}}$ is the impurity concentration and $u$ is
the strength of one single impurity. The Feynman diagram used in the
AG approach is shown in figure~\ref{Fig:FeynmanAG}. In the case of
random chemical potential, one can verify that $A_{2} = 1$. If
long-range correlated disorder \cite{Sarma11} is considered, $A_{2}$
would receive a non-trivial correction. Making use of the fact
$A_{2} = 1$, the self-consistent equations become
\begin{eqnarray}
A_{1} &=& 1 + n_{\mathrm{imp}}u^2 A_{1}\int
\frac{d^2\mathbf{k}}{(2\pi)^{2}}\frac{1}{A_{1}^{2}\omega_{n}^{2} +
v^{2}k^{2} + A_{3}^{2}\Delta^{2}}, \\
A_{3} &=& 1 + n_{\mathrm{imp}}u^2 A_{3}\int
\frac{d^{2}\mathbf{k}}{(2\pi)^{2}}\frac{1}{A_{1}^{2}\omega_{n}^{2} +
v^{2}k^{2}+A_{3}^{2}\Delta^{2}}.
\end{eqnarray}
It is clear that $A_{1}=A_{3}=A$. The gap equation becomes
\begin{eqnarray}
\Delta = 2gT\sum_{\omega_{n}}\int \frac{d^{2}\mathbf{k}}{(2\pi)^{2}}
\frac{A\Delta}{A^{2}\omega_{n}^{2}+v^2k^{2}+A^{2}\Delta^{2}}.
\end{eqnarray}
Carrying out the integration over momenta, we get two coupled
equations
\begin{eqnarray}
A &=& 1+\frac{n_{\mathrm{imp}}u^2}{4\pi v^2}A\ln\left(1 +
\frac{v^2\Lambda^2}{A^{2}\omega_{n}^{2}+A^{2}\Delta^{2}}\right),
\\
\Delta &=& \frac{gT}{2\pi v^2} \sum_{\omega_{n}}A\Delta\ln\left(1 +
\frac{v^2\Lambda^2}{A^{2}\omega^{2}+A^{2}\Delta^{2}}\right).
\end{eqnarray}

\begin{figure}
\center
\includegraphics[width=2.2in]{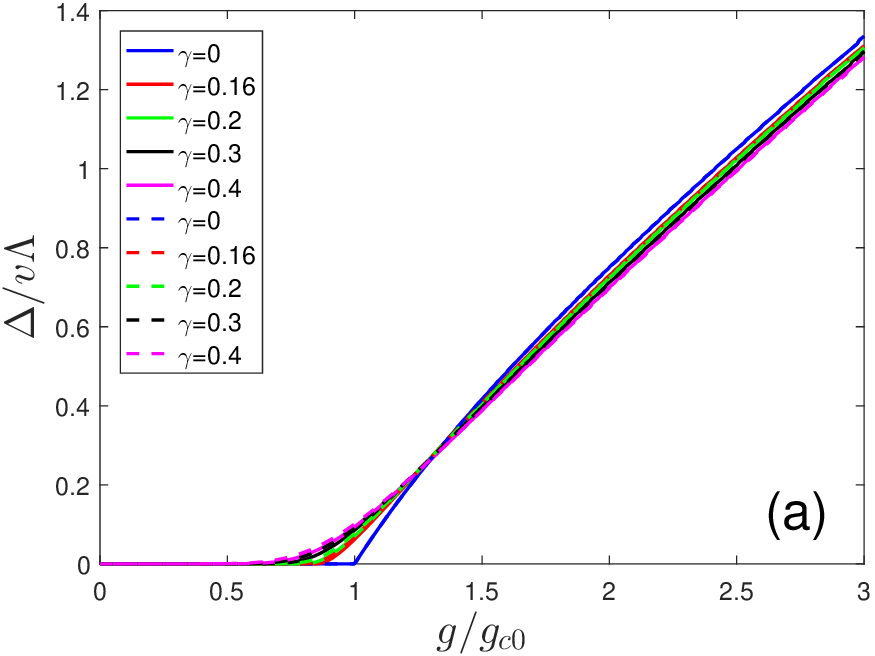}
\includegraphics[width=2.2in]{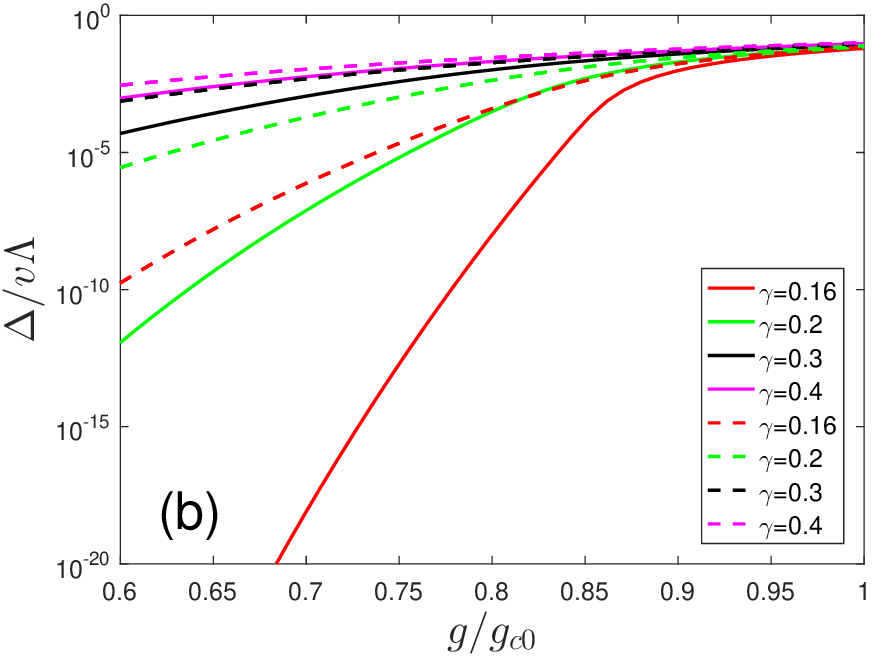}
\caption{Dependence of $\Delta$ on $g$ at different values of
impurity strength $\gamma$ for 2D DSM. Vertex correction is
neglected for solid lines, but incorporated for dashed lines.
\label{Fig:Gap}}
\end{figure}

We now take the zero temperature limit, namely $T = 0$, and then
make the re-scaling transformations
$\frac{\omega}{v\Lambda}\rightarrow\omega$ and
$\frac{\Delta}{v\Lambda}\rightarrow\Delta$, which leads us to
\begin{eqnarray}
A &=& 1+\frac{\gamma}{2} A\ln\left(1 +
\frac{1}{A^{2}\omega^{2}+A^{2}\Delta^{2}}\right),\label{Eq:GapEqAGA}
\\
1 &=& \frac{g}{g_{c0}}\frac{1}{2\pi}\int_{-\infty}^{+\infty} d\omega
A\ln\left(1+\frac{1}{A^{2}\omega^{2}+A^{2}\Delta^{2}}\right),
\label{Eq:GapEqAGB}
\end{eqnarray}
where
\begin{eqnarray}
\gamma = \frac{n_{\mathrm{imp}}u^2}{2\pi v^2}.
\label{Eq:gammaDefinition}
\end{eqnarray}
Upon approaching the QCP, $g \rightarrow g_{c}$, so the above
equations are simplified to
\begin{eqnarray}
A &=& 1+\frac{\gamma}{2}A\ln\left(1 +
\frac{1}{A^{2}\omega^{2}}\right), \label{Eq:AFunZeroGap}
\\
1 &=& \frac{g_{c}}{g_{c0}}\frac{1}{2\pi}
\int_{-\infty}^{+\infty}d\omega A\ln\left(1 +
\frac{1}{A^{2}\omega^{2}}\right).\label{Eq:EqForgc}
\end{eqnarray}

\begin{figure}
\center
\includegraphics[width=2.5in]{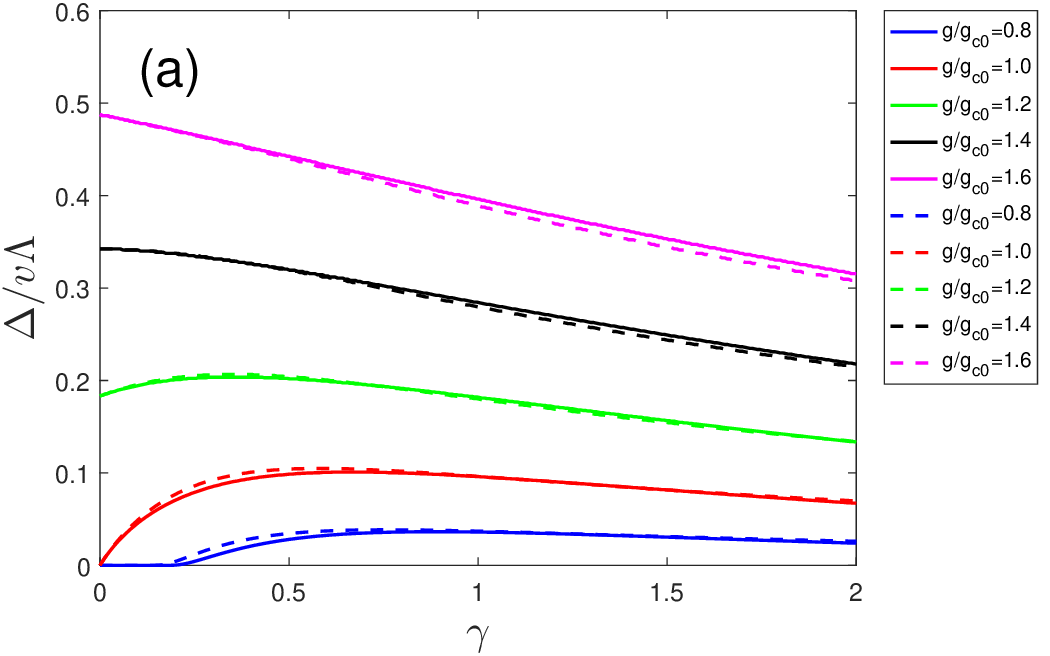}
\includegraphics[width=2.5in]{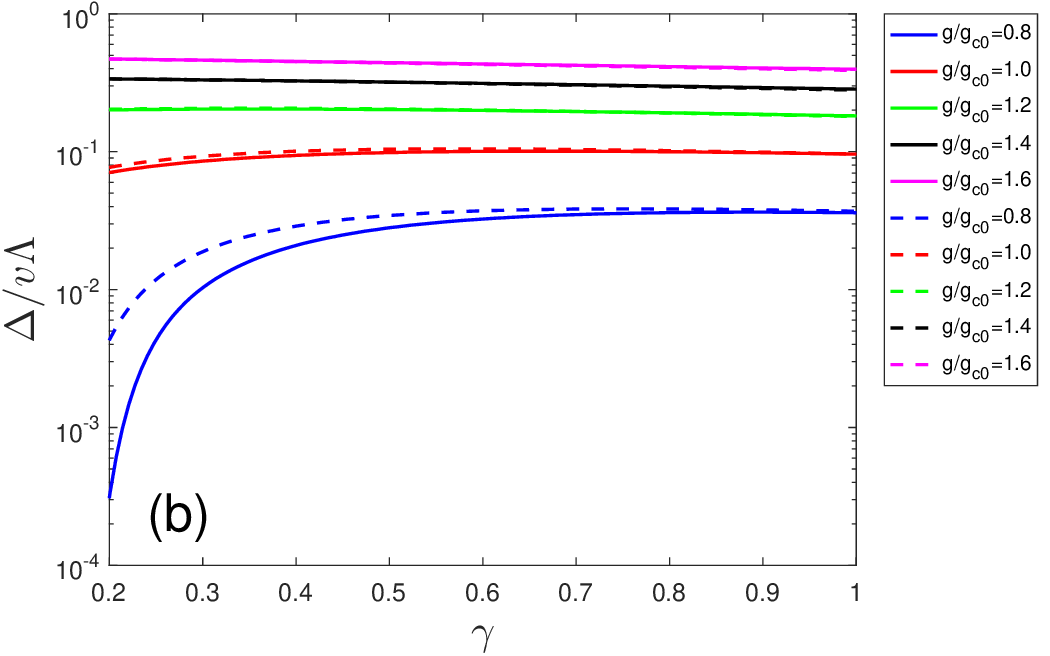}
\caption{Dependence of $\Delta$ on $\gamma$ at different values of
pairing interaction strength $g$ for 2D DSM. Vertex correction is
neglected for solid lines, but incorporated for dashed lines.
\label{Fig:Gapgamma}}
\end{figure}

Numerical results of equation~(\ref{Eq:AFunZeroGap}) are shown in
figures~\ref{Fig:AFun}(a) and (b) by the solid lines. In the zero-energy limit, $\omega A$
approaches to a constant $\Gamma$, which is the disorder scattering
rate. If $\omega$ decreases from the energy scale of $\Gamma$,
$\omega A$ approaches to a constant. Above the scale of $\Gamma$, $A
\rightarrow 1$. The asymptotic behavior of $A$ is approximately
described by
\begin{eqnarray}
A\sim\left\{
\begin{array}{lll}
\frac{\Gamma}{|\omega|} & \mathrm{if} & |\omega|\ll \Gamma,
\\
1 &\mathrm{if} & |\omega|\gg \Gamma.
\end{array}\right.
\end{eqnarray}

\begin{figure}
\center
\includegraphics[width=2.2in]{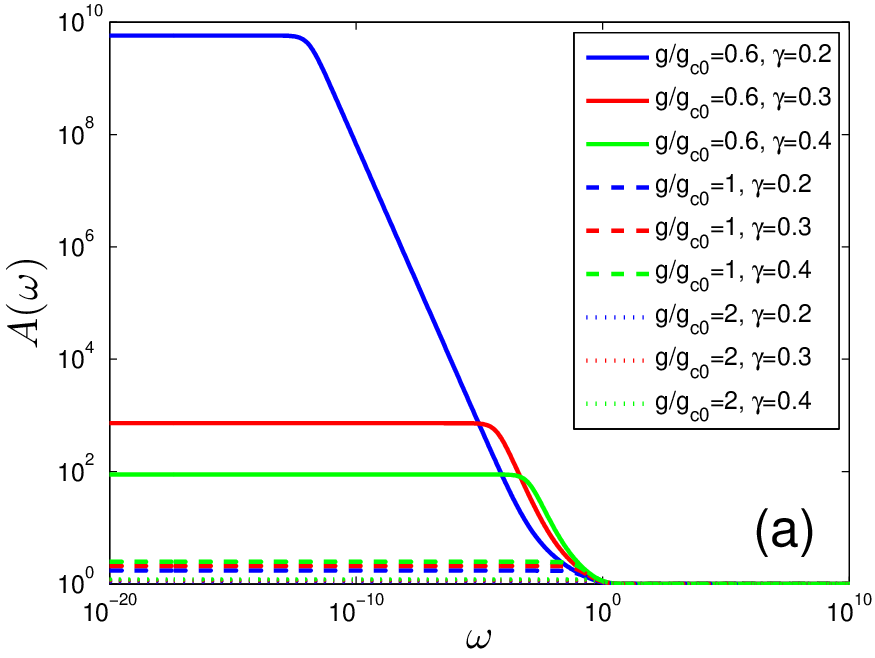}
\includegraphics[width=2.2in]{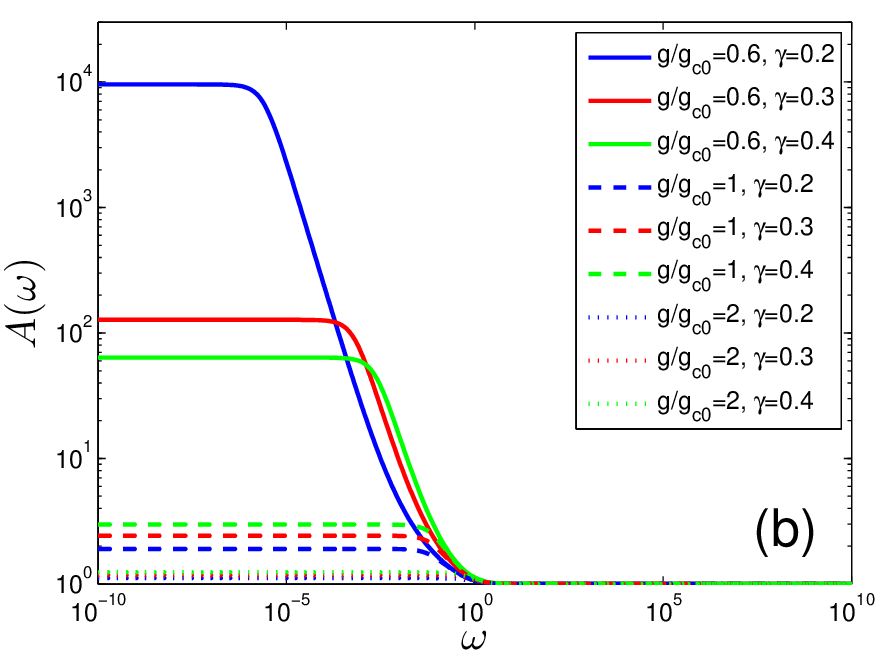}
\caption{Dependence of $A$ on $\omega$ at different values of $g$
and $\gamma$ for 2D DSM. Vertex correction is absent in (a) and
present in (b). \label{Fig:AFunGapPhase}}
\end{figure}

The integral appearing in equation~(\ref{Eq:EqForgc}) is divergent,
which implies that $g_{c}\rightarrow 0$. Therefore, even an
arbitrarily weak attraction suffices to trigger Cooper pairing
instability. In this regard, random chemical potential tends to
promote the formation of Cooper pairing, in agreement with the
result of Nandkishore \emph{et al.} \cite{Nandkishore13}. The
dependence of the gap $\Delta$ on $g$ at different values of
$\gamma$ is displayed in figures~\ref{Fig:Gap}(a) and (b) by the solid lines. We
present the relation between $\Delta$ and $\gamma$ with different
values of $g$ in figures~\ref{Fig:Gapgamma}(a) and (b) by the solid lines. The figures
exhibit that $\Delta$ is enhanced by weak disorder and then
suppressed gradually by strong disorder if the attraction is weak.
However, for sufficiently large attraction, the gap $\Delta$ is
suppressed by disorder monotonously. These results are qualitatively
the same as that obtained by Potirniche \emph{et al.} through using
self-consistent BdG equations \cite{Potirniche14}.

The $\omega$-dependence of $A(\omega)$, obtained from the solutions
of equations~(\ref{Eq:GapEqAGA}) and (\ref{Eq:GapEqAGB}), are shown
in figure~\ref{Fig:AFunGapPhase}(a). The function $A(\omega)$ always
approaches to a finite value in the low energy regime, which is
expected since the gap $\Delta$ provides a cutoff. According to
figure~\ref{Fig:AFunGapPhase}(a), if the pairing interaction is
relatively weak, $A(\omega)$ goes to a larger constant value at low
energies for smaller $\gamma$. If the pairing interaction is
relatively strong, however, $A(\omega)$ takes a larger constant
value at low energies for larger $\gamma$.

\begin{figure}
\center
\includegraphics[width=3.8in]{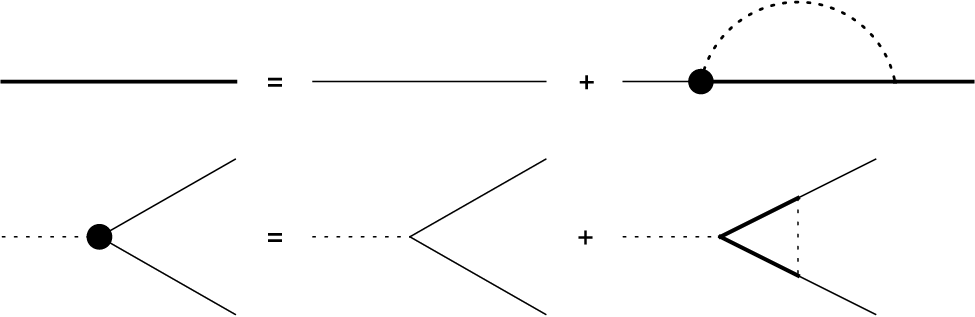}
\caption{Feynman diagram for the fermion self-energy in the presence
of vertex correction. Notice that the dressed fermion propagator is
utilized in the vertex correction. \label{Fig:FeynmanAGBeyond}}
\end{figure}

\subsection{Beyond AG approximation \label{Sec:AGBeyond}}

We emphasize that the gap equation analysis of the disorder effects
based on the original AG formalism is actually problematic. The
validity of AG method relies on a crucial assumption that $k_{F}l
\gg 1$, which is equivalent to $k_F \gg \Gamma$ since $\Gamma \sim
1/l$. This inequality is satisfied in an ordinary metal in which the
Fermi momentum is large and the scattering rate $\Gamma$ can be made
sufficiently small if random chemical potential is supposed to be
weak enough. For a 2D DSM, however, we know that $k_{F} = 0$.
Additionally, even an arbitrarily weak random chemical potential is
able to drive 2D DSM to become a CDM \cite{Ludwig94, Evers08,
Foster12, Fradkin86, Durst00, Ostrovsky06, Katanin13}, which in turn
generates a finite $\Gamma$. Thus, the above inequality certainly
breaks down, and the vertex correction can no longer be regarded as
unimportant.

To examine whether the interesting conclusion reached by employing
the original AG approximation is robust, we need to incorporate the
vertex correction explicitly in the self-consistent gap equations.
The Feynman diagram for fermion self-energy that contains the vertex
correction is shown in figure~\ref{Fig:FeynmanAGBeyond}.

The correction to the fermion-disorder coupling vertex is given by
\begin{eqnarray}
\Xi(\omega,\mathbf{p},\mathbf{q}) = 1 + n_{\mathrm{imp}}u^{2}\int
\frac{d^2\mathbf{k}}{(2\pi)^{2}}G^{F}(\omega,\mathbf{k}+\mathbf{p})
G^{F}(\omega,\mathbf{k}+\mathbf{q}),
\end{eqnarray}
which can be further written as
\begin{eqnarray}
\fl \Xi(\omega,\mathbf{p},\mathbf{q}) &=& 1 + n_{\mathrm{imp}}u^2\int
\frac{d^2\mathbf{k}}{(2\pi)^{2}} \left[-A_{1}^{2}\omega^2 +
v^2\left(\mathbf{k}+\mathbf{p}\right)\cdot \left(\mathbf{k} +
\mathbf{q}\right) + A_{3}^{2}\Delta^2\right]\nonumber
\\
\fl &&\times\left[A_{1}^{2}
\omega^{2} + v^2\left(\mathbf{k}+\mathbf{p}\right)^{2} +
A_{3}^{2}\Delta^{2}\right]^{-1}\left[A_{1}^{2}\omega^{2} +
v^2\left(\mathbf{k}+\mathbf{q}\right)^{2} +
A_{3}^{2}\Delta^{2}\right]^{-1}.
\end{eqnarray}
Using the transformation $\mathbf{k} + \mathbf{q} \rightarrow
\mathbf{k}$, and employing the Feynman parameterization
\begin{eqnarray}
\frac{1}{AB}=\int_{0}^{1}dx\frac{1}{\left[xA+(1-x)B\right]^2},
\end{eqnarray}
we convert the vertex correction into the form
\begin{eqnarray}
\fl \Xi(\omega,\mathbf{p},\mathbf{q}) &=& 1 + n_{\mathrm{imp}}u^2
\int_{0}^{1} dx \int\frac{d^2\mathbf{k}}{(2\pi)^{2}}
\left[-A_{1}^{2}\omega^2 + v^2\left(\mathbf{k} + \mathbf{p} -
\mathbf{q}\right)\cdot \mathbf{k} + A_{3}^{2}\Delta^2\right]
\nonumber \\
\fl &&\times\left[A_{1}^{2}\omega^{2} + v^2
\left(\mathbf{k}+x\left(\mathbf{p}-\mathbf{q}\right)\right)^{2} +
A_{3}^{2}\Delta^{2} + x(1-x)v^2 \left(\mathbf{p} -
\mathbf{q}\right)^{2}\right]^{-2}.
\end{eqnarray}
We further make the replacement $\mathbf{k}+x\left(\mathbf{p} -
\mathbf{q}\right)\rightarrow \mathbf{k}$, and then obtain
\begin{eqnarray}
\fl \Xi(\omega,\mathbf{p},\mathbf{q}) &=&
1+n_{\mathrm{imp}}u^2\int_{0}^{1}dx \int
\frac{d^2\mathbf{k}}{(2\pi)^{2}} \Bigg\{\frac{1}{A_{1}^{2}
\omega^{2} +v^2k^{2}+A_{3}^{2}\Delta^{2} + x(1-x)v^2
\left(\mathbf{p}-\mathbf{q}\right)^{2}}\nonumber
\\
\fl &&-\frac{2A_{1}^{2}\omega^{2}+2x(1-x)v^2(\mathbf{p} -
\mathbf{q})^{2}}{\left[A_{1}^{2}\omega^{2} + v^2k^{2}
+ A_{3}^{2}\Delta^{2} + x(1-x)v^2 \left(\mathbf{p} -
\mathbf{q}\right)^{2}\right]^{2}}\Bigg\}.
\end{eqnarray}
One can verify that $\Xi(\omega,\mathbf{p},\mathbf{q})$ is the
function of $|\mathbf{p}-\mathbf{q}|$, which means
$\Xi(\omega,\mathbf{p},\mathbf{q})\equiv
\Xi(\omega,|\mathbf{p}-\mathbf{q}|)$. Performing the integrations
over $k$ and $x$ leads to the following vertex correction:
\begin{eqnarray}
\fl \Xi(\omega,|\mathbf{p}-\mathbf{q}|) &=& 1+\frac{\gamma}{2}
\left\{\ln\left(\frac{A_{1}^{2}\omega^{2} + v^2\Lambda^{2} +
A_{3}^{2}\Delta^{2}}{A_{1}^{2} \omega^{2} +
A_{3}^{2}\Delta^{2}}\right)\right.\nonumber
\\
\fl &&+\frac{4A_{1}^{2}\omega^{2}+v^2\left|\mathbf{p} -
\mathbf{q}\right|^{2}}{v|\mathbf{p}-\mathbf{q}|\sqrt{J}}
\ln\left(\frac{\sqrt{J}-v\left|\mathbf{p} -
\mathbf{q}\right|}{\sqrt{J} +
v\left|\mathbf{p}-\mathbf{q}\right|}\right)\nonumber
\\
\fl &&\left.-\frac{4A_{1}^{2}\omega^2+v^2\left|\mathbf{p} -
\mathbf{q}\right|^{2}}{v\left|\mathbf{p}-\mathbf{q}\right|
\sqrt{J+4v^2\Lambda^{2}}}\ln\left( \frac{\sqrt{J+4v^2\Lambda^{2}} -
v\left|\mathbf{p}-\mathbf{q}\right|}{ \sqrt{J+4v^2\Lambda^{2}} +
v\left|\mathbf{p}-\mathbf{q}\right|}\right)\right\},
\end{eqnarray}
where $\gamma$ is defined in equation~(\ref{Eq:gammaDefinition}), and
\begin{eqnarray}
J = 4\left(A_{1}^{2}\omega^{2} + A_{3}^{2}
\Delta^{2}\right) + v^2
\left|\mathbf{p}-\mathbf{q}\right|^{2}.
\end{eqnarray}

After including the vortex correction, the equations for $A_{1}$ and
$A_{3}$ now become
\begin{eqnarray}
A_{1} &=& 1 + n_{\mathrm{imp}}u^2A_{1}\int
\frac{d^2\mathbf{k}}{(2\pi)^{2}}
\frac{\Xi(\omega,k)}{A_{1}^{2}\omega^{2} + v^{2}k^{2} +
A_{3}^{2}\Delta^{2}} \\
A_{3} &=& 1+n_{\mathrm{imp}}u^2 A_{3} \int
\frac{d^{2}\mathbf{k}}{(2\pi)^{2}}
\frac{\Xi(\omega,k)}{A_{1}^{2}\omega^{2}+v^{2}k^{2} +
A_{3}^{2}\Delta^{2}}.
\end{eqnarray}
We now employ the re-scaling transformations
$\frac{k}{\Lambda}\rightarrow k$,
$\frac{\omega}{v\Lambda}\rightarrow\omega$, and
$\frac{\Delta}{v\Lambda}\rightarrow\Delta$, and making use of the
relation $A_{1} = A_{3} = A$, obtain
\begin{eqnarray}
A = 1 + \gamma A\int_{0}^{1} dkk\frac{\Xi(\omega,k)}
{A^{2}\omega^{2} + k^{2} + A^{2}\Delta^{2}},
\end{eqnarray}
where the gap $\Delta$ is still given by equation~(\ref{Eq:GapEqAGB}).

In the limit of $\Delta = 0$, the system stays in the non-SC phase.
We present the $\omega$-dependence of $A$ and $\omega A$ in
figure~\ref{Fig:AFun}(a) and figure~\ref{Fig:AFun}(b) respectively by the dashed
lines. Comparing the solid curves and dashed curves,
we can observe that the disorder scattering rate $\Gamma$ is made
larger by the inclusion of the vertex correction. Katanin
\cite{Katanin13} performed a functional RG analysis of this problem,
and also found a larger scattering rate $\Gamma$ and a larger
$\rho(0)$ comparing to those obtained by using the self-consistent
Born approximation (SCBA). Sinner and Ziegler \cite{Sinner17}
reported a $1/N$-expansion study, which claims that higher order
corrections lead to enhancement of disorder scattering rate and DOS.

Now we consider the SC phase where $\Delta \neq 0$. The
$g$-dependence of gap $\Delta$ is displayed in figures~\ref{Fig:Gap}
(a) and (b) by the dashed lines. A clear result is that the zero-energy gap $\Delta$ is
increased when the vertex correction is taken into account, which
means that including vertex correction leads to further promotion of
superconductivity. Dependence of $\Delta$ on $\gamma$ with different
values of $g$ is presented in figures~\ref{Fig:Gapgamma}(a) and (b) by the
dashed lines. We can find that the qualitative property of the disorder effect on
superconductivity remains nearly the same after including the vertex
correction.

\begin{figure}
\center
\includegraphics[width=2.2in]{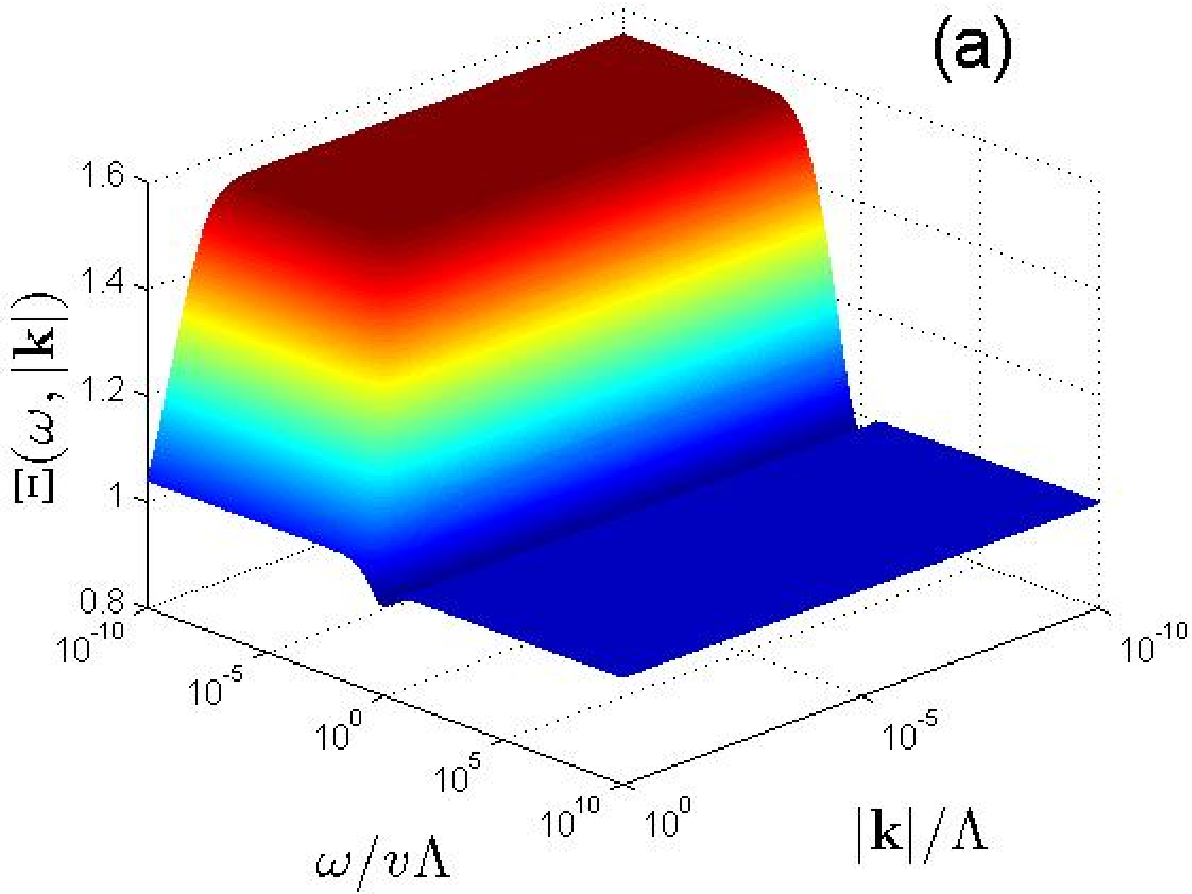}
\includegraphics[width=2.2in]{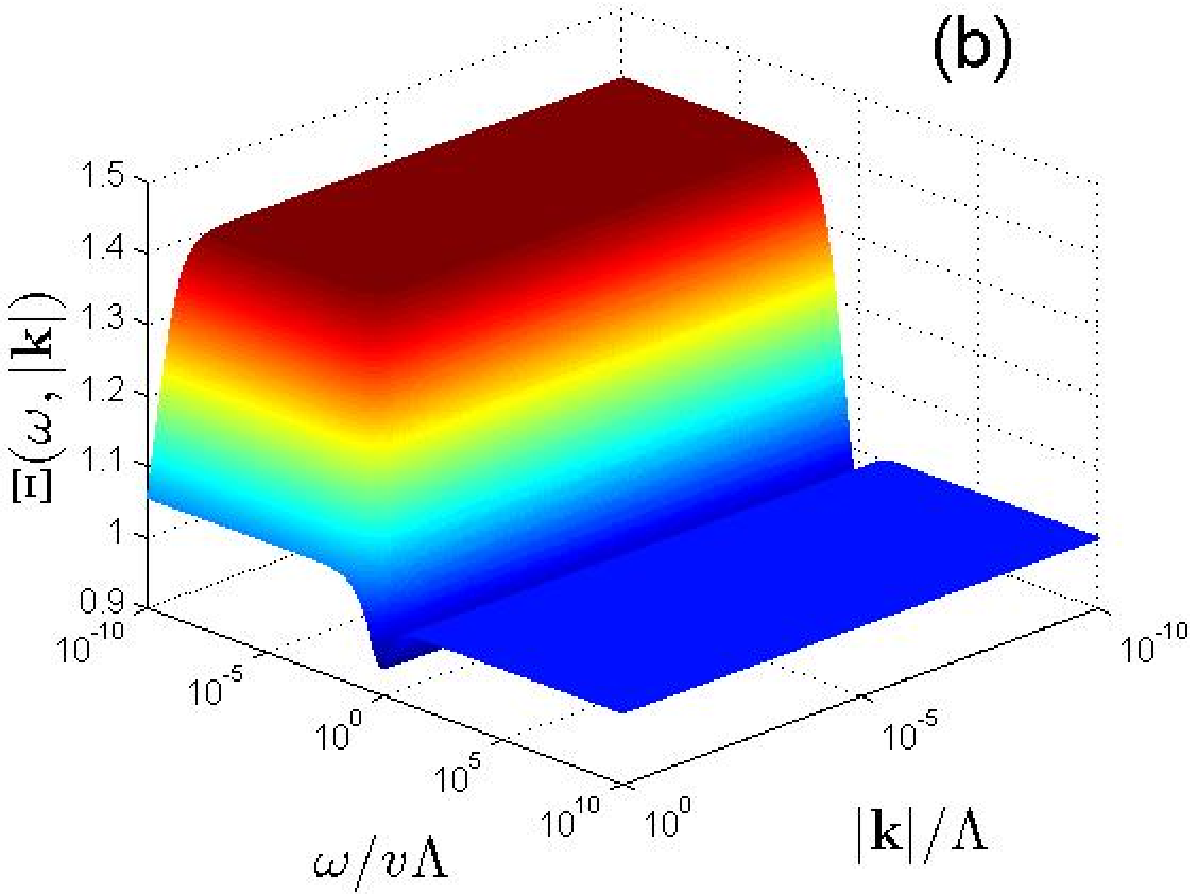}
\includegraphics[width=2.2in]{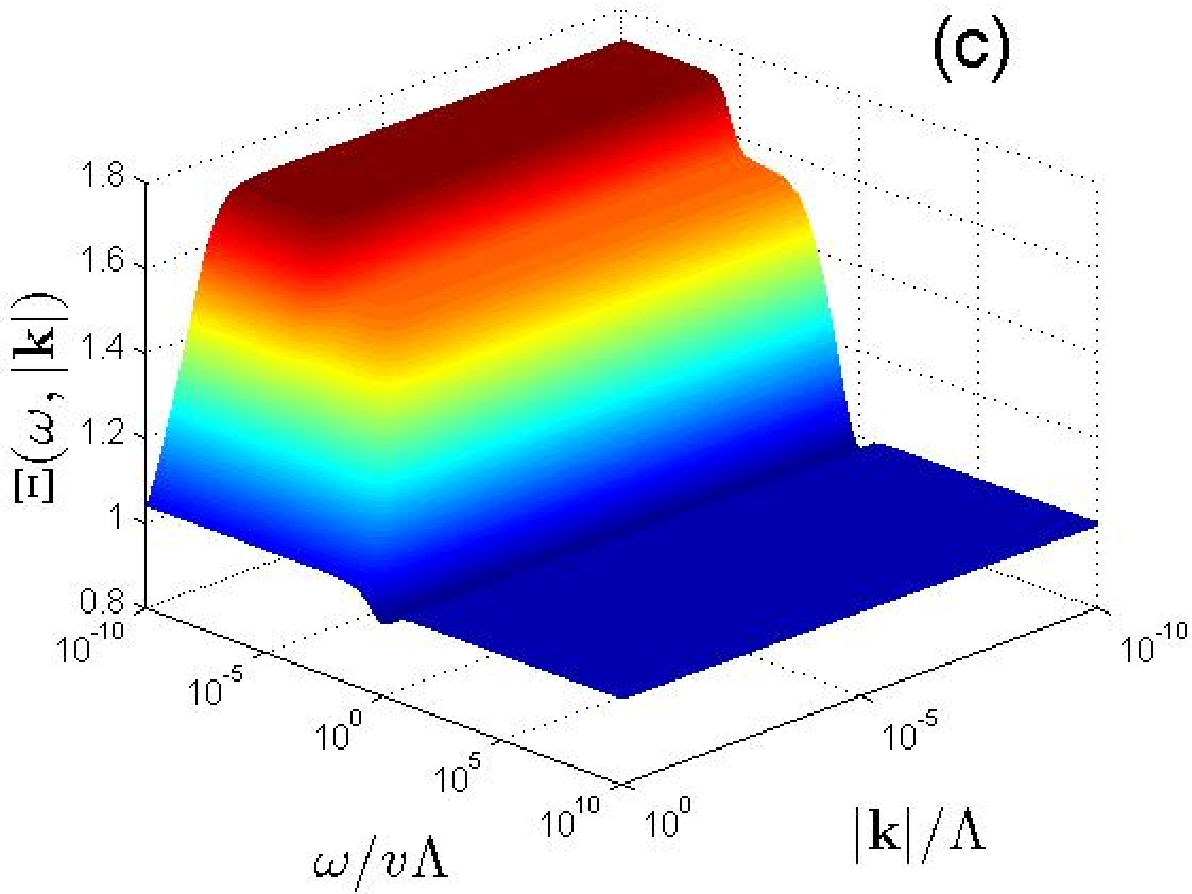}
\includegraphics[width=2.2in]{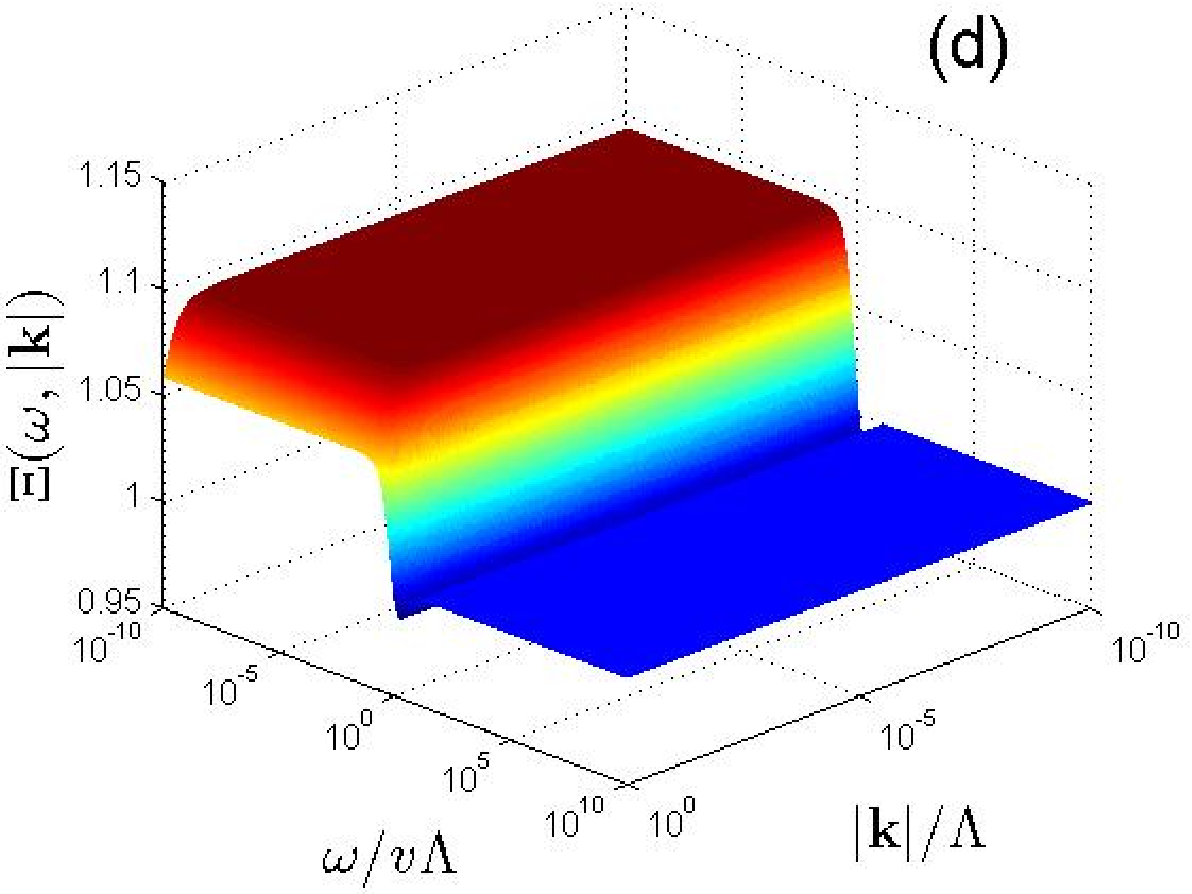}
\caption{(a) Dependence of the vertex function
$\Xi(\omega,|\mathbf{k}|)$ on energy and momentum with $\gamma =
0.2$ and $\gamma = 0.4$ in (a) and (b) respectively. $\Delta = 0$ is
taken in $(a)$ and (b). Dependence of $\Xi(\omega,|\mathbf{k}|)$ on
energy and momentum with $g = 0.6$ and $g = 2$ in (c) and (d)
respectively. $\gamma=0.2$ is taken in (c) and (d).
\label{Fig:Vertex}}
\end{figure}

In the case of $\Delta \neq 0$, we show the dependence of
$A(\omega)$ on $\omega$ in the presence of vertex correction in
figure~\ref{Fig:AFunGapPhase}(b). It is obvious that $A(\omega)$
approaches to a smaller constant comparing to the one shown in
figure~\ref{Fig:AFunGapPhase}(a). The reason for this behavior is
that the gap $\Delta$ becomes larger after including the vertex
correction and thus leads to a stronger suppression effect for the
disorder scattering rate.

The energy and momentum dependence of the vertex function
$\Xi(\omega, k)$ are shown in figure~\ref{Fig:Vertex}. An apparent
fact is that the vertex correction is important at low energies and
small momenta, and can be nearly ignored only when the energy and
momentum are sufficiently large. As the pairing interaction gets
stronger, the vertex correction becomes less important
\cite{Nandkishore13}.

The SM-SC QCP exists in a clean 2D DSM, but is eliminated when the
system contains weak random chemical potential. Since there is
always certain amount of impurity in the material, it seems
extremely difficult to realize and probe the predicted quantum
critical phenomena at such a QCP.

\section{Superconductivity in 3D Dirac semimetal \label{Sec:3DDSM}}

In this section, we will investigate the fate of superconductivity
formed by Cooper pairing of 3D Dirac fermions, which could emerge at
low energies at the QCP between a normal band insulator and a 3D
topological insulator. This type of 3D DSM has been observed in
TiBiSe$_{2-x}$S$_{x}$ \cite{XuSuYang11, Sato11} and
Bi$_{2-x}$In$_{x}$Se$_{3}$ \cite{Brahlek12, Wu13} by fine tuning the
doping level. Theoretical works \cite{WangZhiJun12, Wangzhijun13}
predicted that a crystal-symmetry protected 3D DSM might be realized
in such materials as A$_{3}$Bi (A=Na, K, Rb) and Cd$_{3}$As$_{2}$.
Shortly after this prediction, ARPES and quantum transport
measurements have reported evidence of 3D DSM state in Na$_{3}$Bi
and Cd$_{3}$As$_{2}$ \cite{LiuZK14A, Neupane14, LiuZK14B, Borisenko14,
HeLP14}.

Recent RG analysis revealed that weak attraction is irrelevant in 3D
DSM, and that only sufficiently strong attraction can induce
superconductivity \cite{Roy16} and there is also a QCP separating
the SM and SC phases. In the non-SC phase, the physical effect
caused by the random chemical potential is a subject of considerable
interest \cite{Goswami11, Roy14, Syzranov18Review, Fu17, Sbierski17,
Ominato15, Nandkishore14, PixleySeries}. Recent studies based on
SCBA, RG analysis, and exact numerical simulation all found that
there is a QCP between the SM and CDM phases by adjusting the
strength of random chemical potential \cite{Goswami11, Roy14,
Syzranov18Review, Fu17, Sbierski17, Ominato15}. If the rare region
effect is considered, it was found that arbitrarily weak disorder
drives the system to enter into the CDM phase \cite{Nandkishore14,
PixleySeries}. In this paper, we do not consider the rare region
effect, and focus on the fate of $s$-wave superconductivity.

Similar to 2D DSM, the 3D DSM hosts only discrete Fermi points, and
thus the vertex correction may also be important. In addition, the
zero-energy DOS vanishes in both cases, and might become finite if
the system is turned by random chemical potential into a CDM. We now
parallel the analysis performed in the last section, and include
explicitly the vertex correction in the self-consistent equations
for $A$ and $\Delta$. The disorder effect on superconductivity can
be analogously analyzed.

\subsection{Clean case}

The mean-field Hamiltonian for 3D DSM is formally similar to that of
2D DSM, and will be not explicitly given here. We directly write
down the the gap equation obtained in the clean limit:
\begin{eqnarray}
\Delta = 2g\int\frac{d\omega}{2\pi} \int
\frac{d^{3}\mathbf{k}}{(2\pi)^{3}}
\frac{\Delta}{\omega^{2}+v^2k^{2}+\Delta^{2}},
\end{eqnarray}
where $g$ is the strength parameter for pairing interaction.
Integrating over momenta results in
\begin{eqnarray}
1 &=& \frac{g}{2\pi^{3}v^{2}}\int_{-\infty}^{+\infty}d\omega
\left[\Lambda-\frac{1}{v} \sqrt{\omega^2+\Delta^{2}}
\arctan\left(\frac{v\Lambda}{\sqrt{\omega^2 +
\Delta^{2}}}\right)\right],
\end{eqnarray}
where $\Lambda$ is the momentum cutoff. The critical attraction
strength can be determined by taking $\Delta = 0$, satisfying the
following equation:
\begin{eqnarray}
1 &=& \frac{g_{c0}}{2\pi^{3}v^{2}}\int_{-\infty}^{+\infty}d\omega
\left[\Lambda-\frac{|\omega|}{v}
\arctan\left(\frac{v\Lambda}{|\omega|}\right)\right]
= \frac{g_{c0}\Lambda^{2}}{4\pi^{2}v}.
\end{eqnarray}
The critical value is $g_{c0} = 4\pi^{2}v/\Lambda^{2}$.

\subsection{Analysis by AG method without vertex correction}

Under the original AG approximation, the self-consistent equations
for $A$ and $\Delta$ are given by
\begin{eqnarray}
\fl A &=& 1 + \gamma A\left[1- \sqrt{A^{2}\omega^2 +
A^{2}\Delta^{2}}\arctan\left(\frac{1}{\sqrt{A^{2}\omega^2 +
A^{2}\Delta^{2}}}\right)\right], \label{Eq:AFun3DDirac}
\\
\fl 1 &=& \frac{g}{g_{c0}}\frac{2}{\pi}\int_{-\infty}^{+\infty}
d\omega A \left[1-\sqrt{A^{2}\omega^2 + A^{2}\Delta^{2}}
\arctan\left(\frac{1}{\sqrt{A^{2}\omega^2 +
A^{2}\Delta^{2}}}\right)\right], \label{Eq:Gap3DDirac}
\end{eqnarray}
where
\begin{eqnarray}
\gamma=\frac{n_{\mathrm{imp}}u^2\Lambda}{2\pi^{2} v^2}.
\end{eqnarray}
In the derivation, we have employed the transformations:
$\frac{k}{\Lambda}\rightarrow k$,
$\frac{\omega}{v\Lambda}\rightarrow \omega$, and
$\frac{\Delta}{v\Lambda}\rightarrow\Delta$.

Before analyzing the disorder effect on superconductivity, it is
necessary to first consider the impact of weak disorder on the
low-energy behavior of Dirac fermions in the non-SC phase. This is
an interesting problem and has been studied recently by means of
several different approaches \cite{Goswami11, Roy14,
Syzranov18Review, Fu17, Sbierski17, Ominato15, Nandkishore14,
PixleySeries}.

\begin{figure}
\center
\includegraphics[width=2.2in]{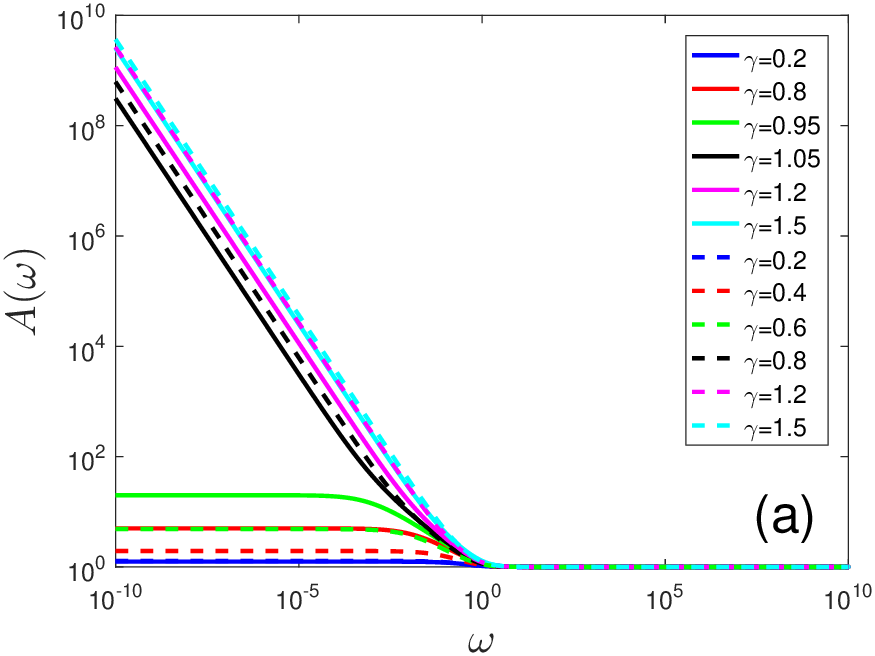}
\includegraphics[width=2.2in]{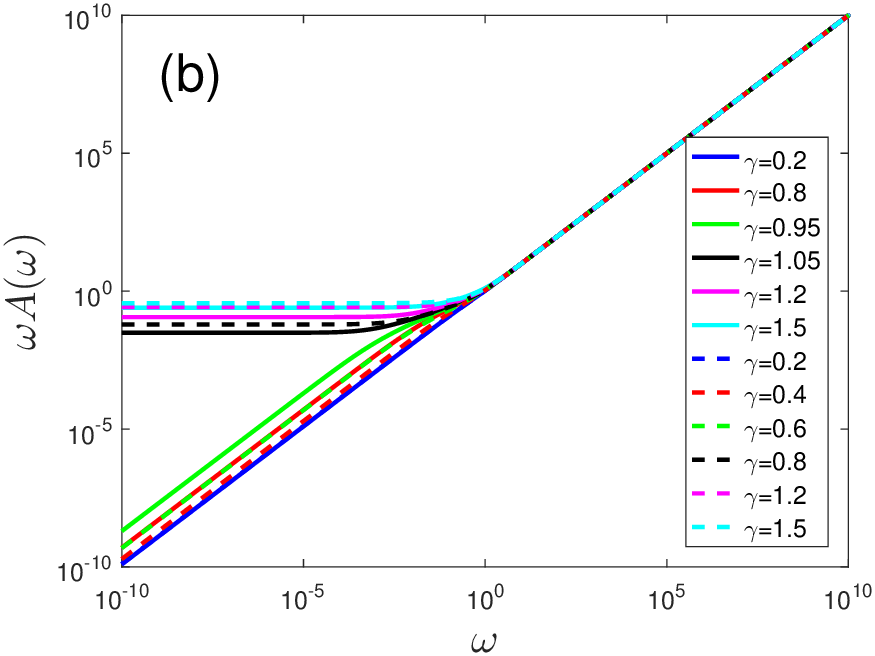}
\caption{(a) Dependence of $A$ and (b) $\omega A$ on $\omega$
with different values of $\gamma$ for 3D DSM. Vertex correction is
neglected for solid lines, but incorporated for dashed lines.
$\Delta=0$ is taken. \label{Fig:AFun3DDirac}}
\end{figure}

\begin{figure}
\center
\includegraphics[width=2.2in]{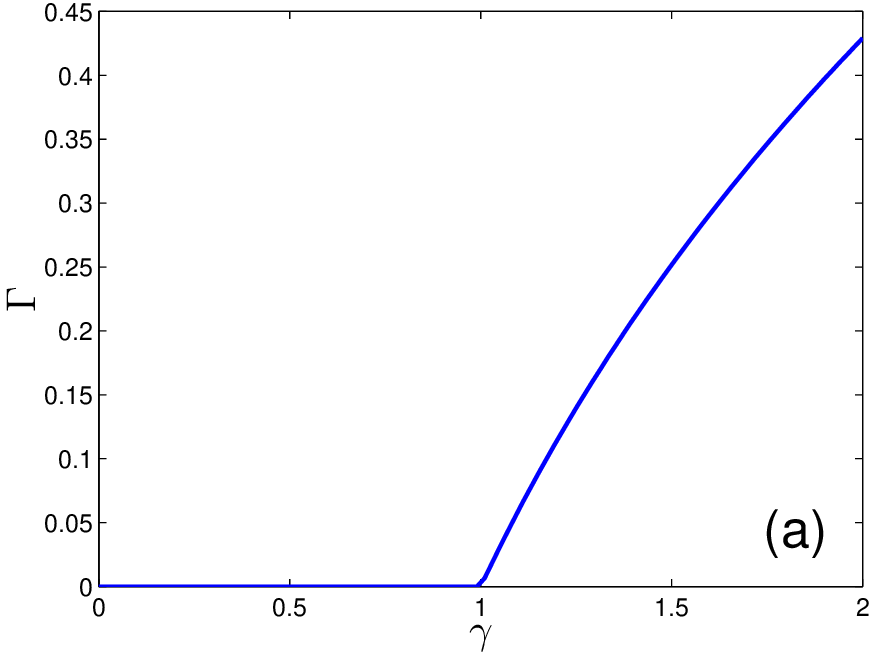}
\includegraphics[width=2.2in]{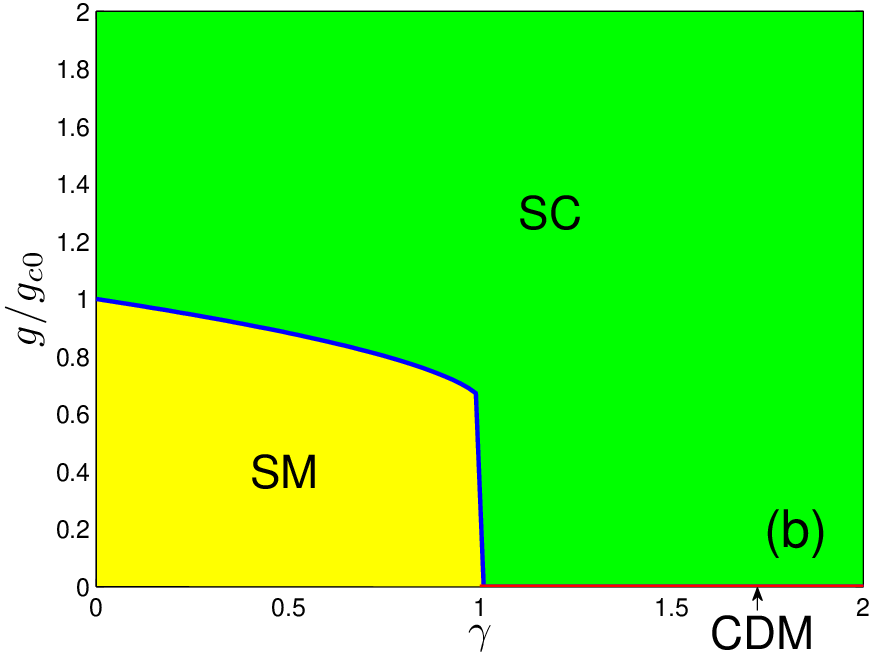}
\includegraphics[width=2.2in]{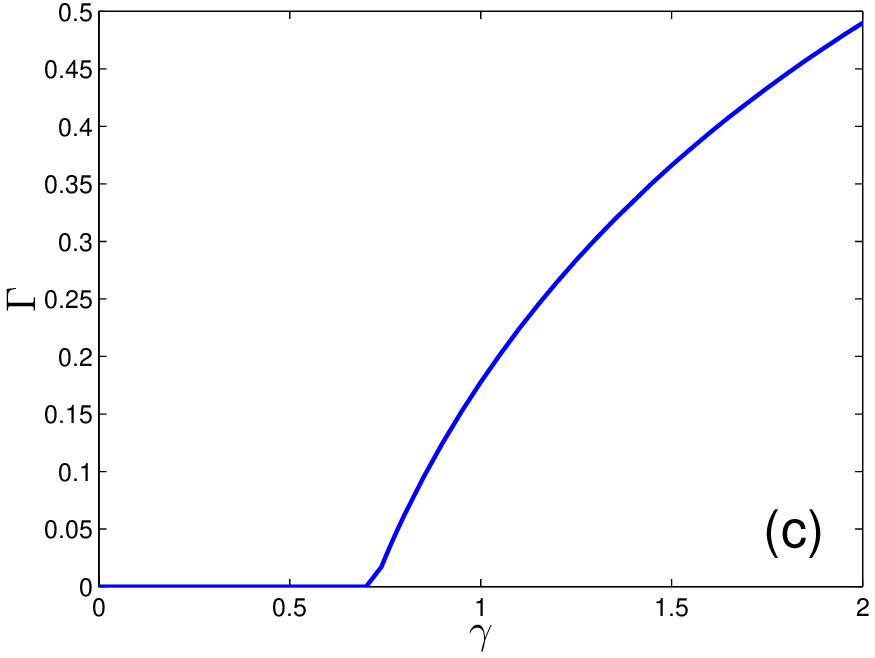}
\includegraphics[width=2.2in]{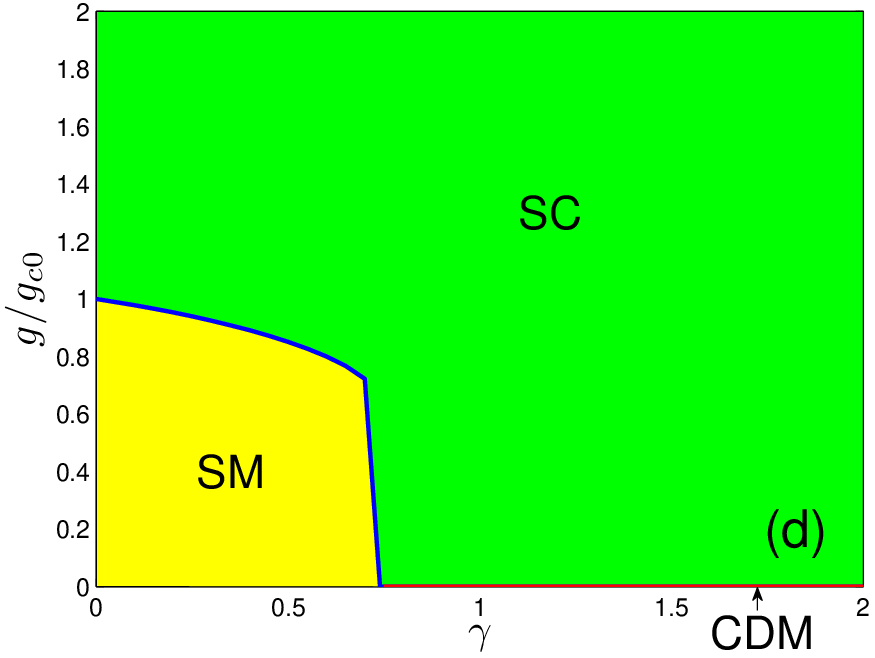}
\caption{(a, c) Dependence of $\Gamma$ on $\gamma$ for 3D DSM where
$\Gamma = \lim_{\omega\rightarrow 0}A(\omega)\omega$. (b, d) Phase
diagram of 3D DSM on the $g$-$\gamma$ plane. Vertex correction is
absent in (a) and (b), and present in (c) and (d). The
SC region is broadened after including the vertex correction.
\label{Fig:3DPhaseDiagram}}
\end{figure}

In the limit of $\Delta = 0$, the equation for $A$ has the form
\begin{eqnarray}
A = 1 + \gamma A\left[1 - A|\omega|
\arctan\left(\frac{1}{A|\omega|}\right)\right].
\label{Eq:AFun3DDiracNoGap}
\end{eqnarray}
The solutions for this equation are shown in
figures~\ref{Fig:AFun3DDirac} (a) and (b) by the solid lines. We find that $A(\omega)$
approaches a finite value in the limit $\omega\rightarrow 0$ if
$\gamma$ is smaller than a critical value $\gamma_{c}$. In contrast,
if $\gamma > \gamma_{c}$, $A(\omega)$ is divergent in the limit
$\omega \rightarrow 0$, yet satisfying
\begin{eqnarray}
\lim \omega A(\omega)\rightarrow \Gamma,
\end{eqnarray}
where $\Gamma$ takes a finite value. The constant $\Gamma$ should be
identified as the disorder scattering rate. A finite DOS is
generated at the Fermi level, which tends to favor
superconductivity. The dependence of $\Gamma$ on $\gamma$ is
depicted in figure~\ref{Fig:3DPhaseDiagram}(a), which clearly shows
that $\gamma_{c} = 1$. Making use of the approximation $\omega
A(\omega)\rightarrow\Gamma$, we rewrite
equation~(\ref{Eq:AFun3DDiracNoGap}) in the form
\begin{eqnarray}
1=\gamma\left[1-\Gamma\arctan\left(\frac{1}{\Gamma}\right)\right].
\end{eqnarray}
By setting $\Gamma = 0$, we find that $\gamma_{c} = 1$. According to
the results presented in figures~\ref{Fig:AFun3DDirac}(a) and (b), and
figures~\ref{Fig:3DPhaseDiagram}(a) and (b), the system undergoes a
quantum phase transition from the SM phase to a CDM phase
at $\gamma = \gamma_{c}$. This result is consistent with previous
studies based on perturbative RG \cite{Goswami11, Roy14}, functional
RG \cite{Sbierski17}, and direct numerical calculation \cite{Fu17}.
We point out here that we do not consider the effects of rare
region for simplicity \cite{Nandkishore13, Nandkishore14,
PixleySeries}.

\begin{figure}
\center
\includegraphics[width=2.2in]{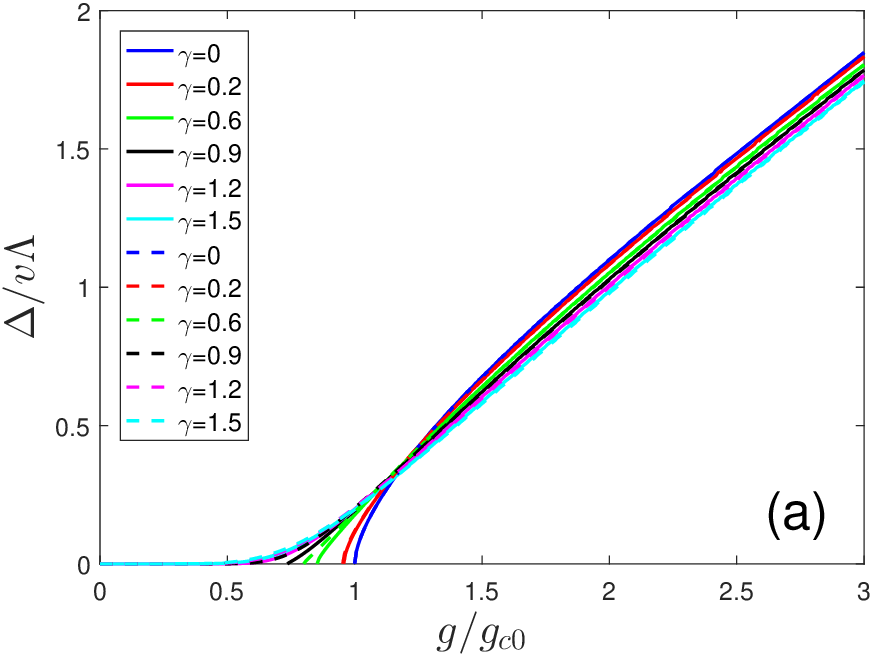}
\includegraphics[width=2.6in]{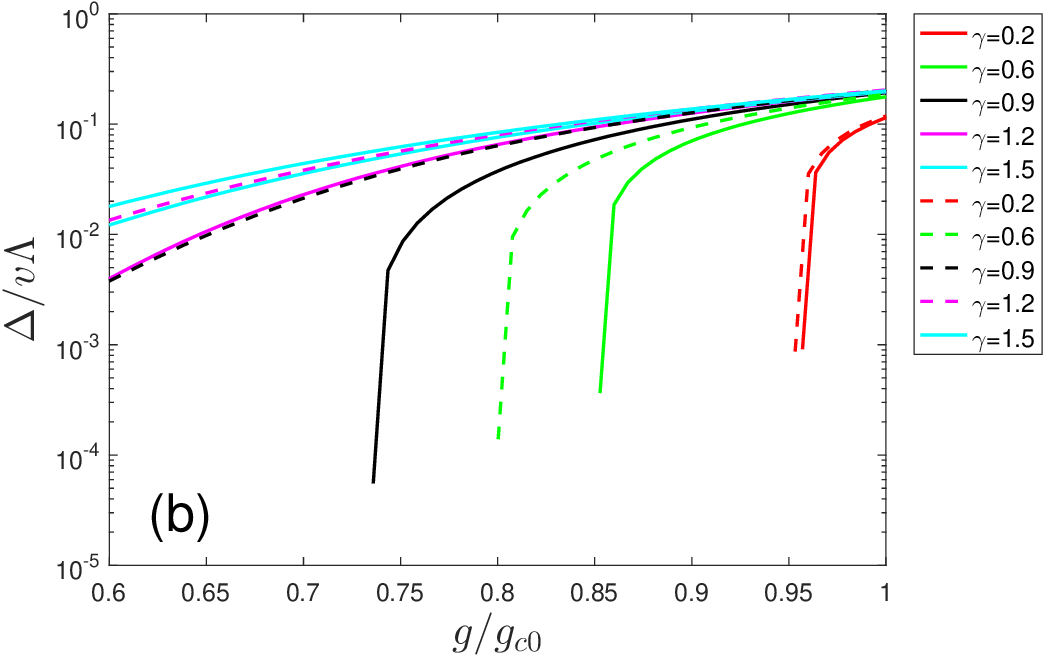}
\caption{Dependence of $\Delta$ on $g$ at different values of
$\gamma$ for 3D DSM. Vertex correction is
neglected for solid lines, but incorporated for dashed lines. \label{Fig:Gap3D}}
\end{figure}

\begin{figure}
\center
\includegraphics[width=2.6in]{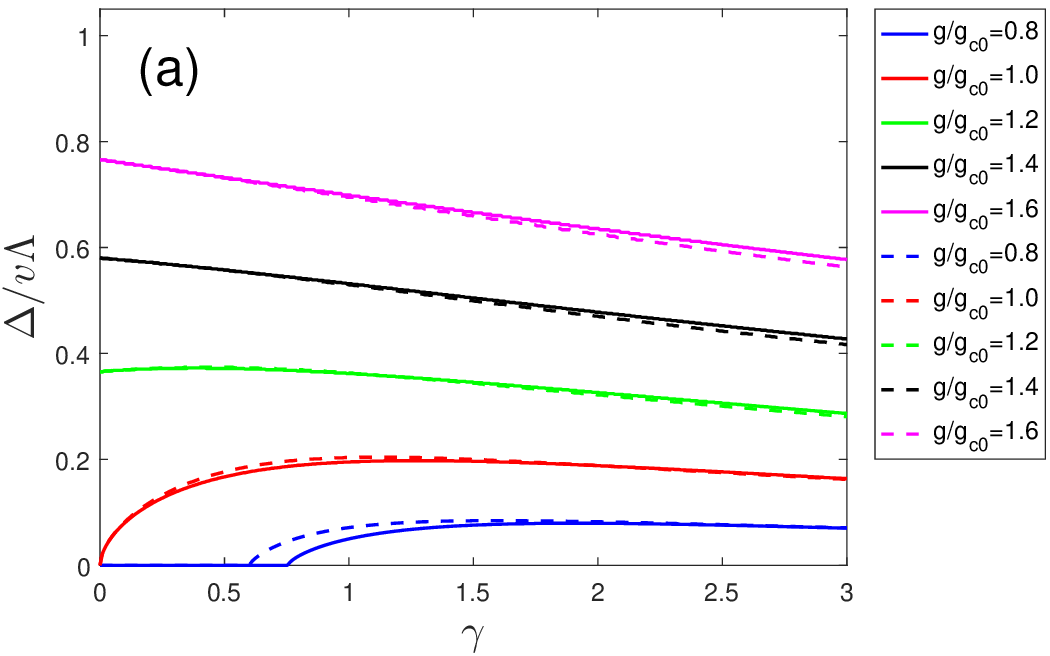}
\includegraphics[width=2.6in]{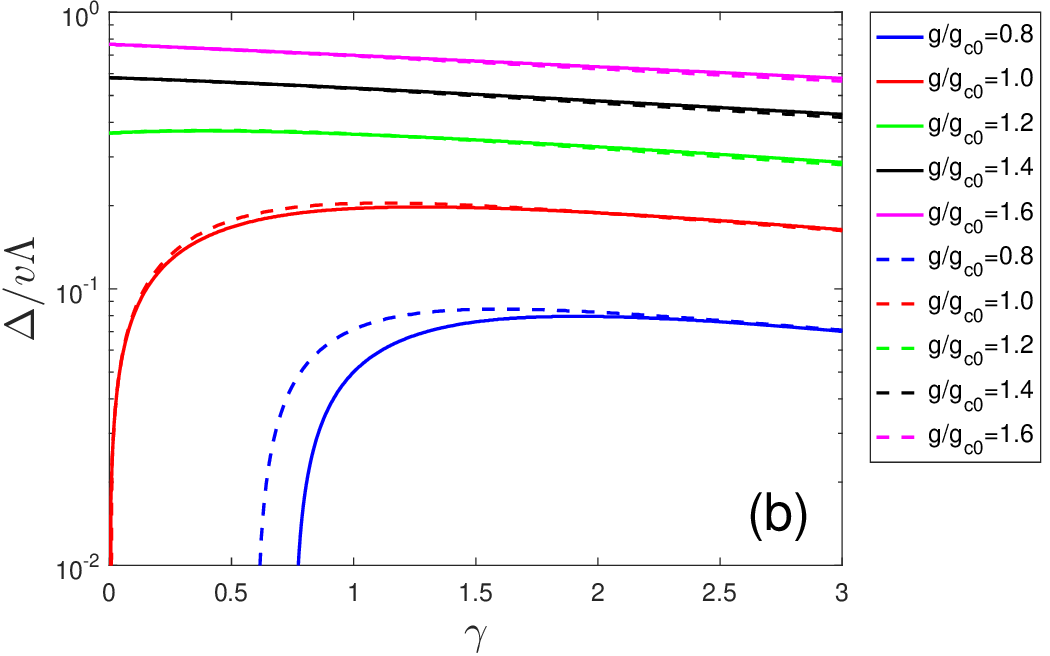}
\caption{Dependence of $\Delta$ on $\gamma$ at different values of
$g$ for 3D DSM. Vertex correction is
neglected for solid lines, but incorporated for dashed lines. \label{Fig:Gapgamma3D}}
\end{figure}

We then turn to solve the coupled equations (\ref{Eq:AFun3DDirac})
and (\ref{Eq:Gap3DDirac}), which will be used to analyze the impact
of disorder on superconductivity. As can be seen from
figure~\ref{Fig:3DPhaseDiagram}(b), the critical value $g_{c}$
decreases as the disorder parameter $\gamma$ grows, indicating that
superconductivity is promoted. More concretely, in the range $0 <
\gamma < 1$, $g_{c}$ is made smaller than $g_{c0}$ but remains
finite. Thus, there is still a SC QCP and it is possible to observe
the corresponding quantum critical phenomena. If $\gamma > 1$, we
take the limit $\Delta \rightarrow 0$ for
equation~(\ref{Eq:Gap3DDirac}) and then obtain an equation for
$g_{c}$:
\begin{eqnarray}
1 = \frac{g_{c}}{g_{c0}}\frac{2}{\pi}\int_{-\infty}^{+\infty}
d\omega A \left[1 - A|\omega|
\arctan\left(\frac{1}{A|\omega|}\right)\right].
\label{Eq:gc3DDiracCDM}
\end{eqnarray}
At low energies, $A(\omega)$ behaves as $A(\omega) \sim
\frac{\Gamma}{|\omega|}$. Accordingly, the integral in
equation~(\ref{Eq:gc3DDiracCDM}) is divergent, indicating that the
critical value of attraction vanishes, i.e., $g_{c} \rightarrow 0$.
It turns out that even arbitrarily weak attraction suffices to form
Cooper pairs. The dependence of $\Delta$ on $g$ at different values
of $\gamma$ can be found in figures~\ref{Fig:Gap3D}(a) and (b) by the
solid lines. The
dependence of $\Delta$ on $\gamma$ at different values of $g$ is
shown in figures~\ref{Fig:Gapgamma3D}(a) and (b) by the solid lines.
We observe that the gap $\Delta$ displays a non-monotonic dependence on $\gamma$ if $g$
is relatively small: $\Delta$ is enhanced by weak disorder but gets
suppressed by sufficiently strong disorder. However, the gap is
always suppressed when $g$ becomes relatively large. This behavior
is qualitatively similar to 2D DSM.

\begin{figure}
\center
\includegraphics[width=2.2in]{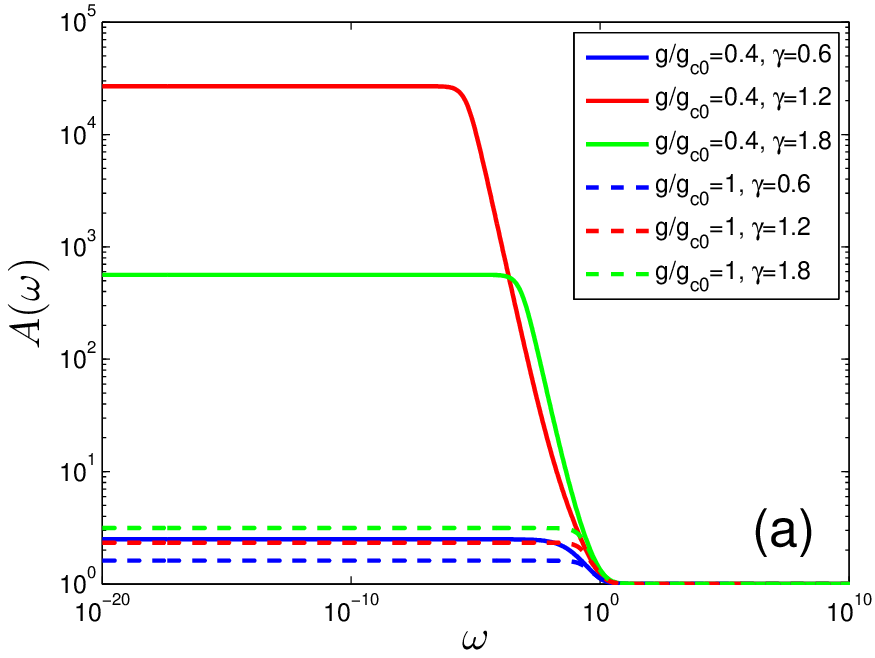}
\includegraphics[width=2.2in]{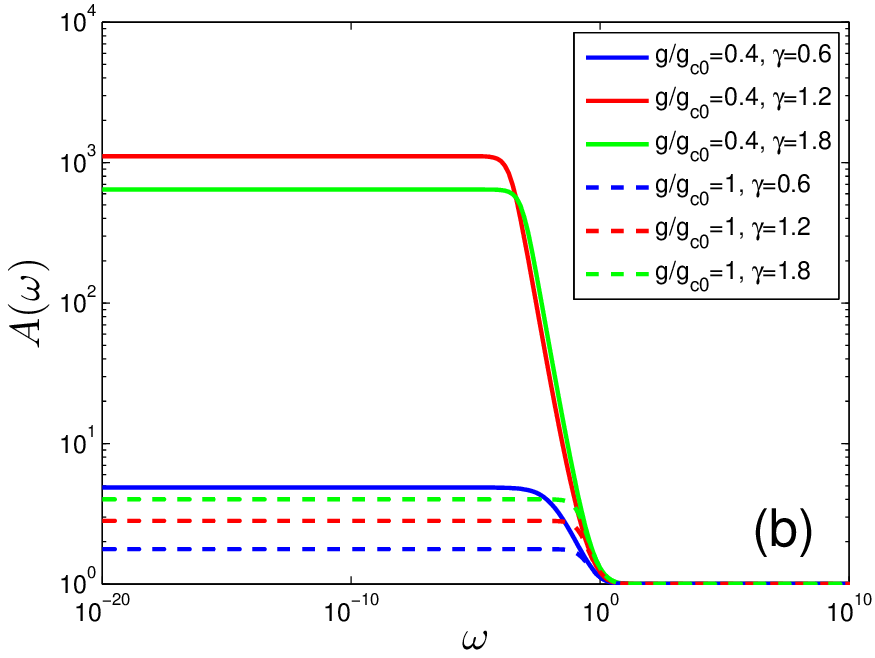}
\caption{Dependence of $A$ on $\omega$ at different values of $g$
and $\gamma$ for 3D DSM. Vertex correction is absent in (a) and
present in (b). \label{Fig:AFunGapPhase3DDirac}}
\end{figure}

The asymptotic behavior of $A(\omega)$ for different values of $g$
and $\gamma$ is shown in figure~\ref{Fig:AFunGapPhase3DDirac}(a). We
can find that $A(\omega)$ generally approach to some finite value
determined by $g$ and $\gamma$. In the SM phase, $\Delta$ is
vanishing, and $A(\omega)$ is saturated to finite value and $\omega
A(\omega)$ vanishes in the lowest energy limit. In the SC phase, the
nonzero gap $\Delta$ serves as a cutoff and prevents $A(\omega)$
from being divergent in the lowest energy limit.

\subsection{Beyond AG approximation}

Paralleling the analysis carried out in the case of 2D DSM, we now
examine the role played by the vertex correction. After including
the vertex correction into the equations of $A$ and $\Delta$,
equation~(\ref{Eq:AFun3DDirac}) becomes
\begin{eqnarray}
A = 1 + \gamma A\int_{0}^{1}dk
\frac{k^{2}\Xi(\omega,k)}{A\omega^{2}+k^{2}+A\Delta^{2}},
\label{Eq:AFun3DDiracVertex}
\end{eqnarray}
but the gap equation~(\ref{Eq:Gap3DDirac}) is not changed.
Straightforward calculations lead us to the following expression for
the vertex function $\Xi$:
\begin{eqnarray}
\fl && \Xi(\omega,\mathbf{p},\mathbf{q}) \equiv
\Xi\left(\omega,|\mathbf{p}-\mathbf{q}|\right) \nonumber \\
\fl &=& 1+\gamma\left[2+2\frac{A^{2}\Delta^{2}+1}{v|\mathbf{p} -
\mathbf{q}|\sqrt{J_{1}}}\ln\left|\frac{\sqrt{J_{1}} -
v\left|\mathbf{p} - \mathbf{q}\right|}{\sqrt{J_{1}} +
v\left|\mathbf{p} -
\mathbf{q}\right|}\right| -\int_{0}^{1}dx\frac{J_{3}}{\sqrt{J_{2}}}
\arctan\left(\frac{1}{\sqrt{J_{2}}}\right)\right],
\label{Eq:VertexCorrection3DDirac}
\end{eqnarray}
where
\begin{eqnarray}
J_{1} &=& 4\left(A^{2}\omega^{2} + 1 + A^{2} \Delta^{2}\right) +
v^{2} \left(\mathbf{p}-\mathbf{q}\right)^{2}, \\
J_{2} &=& A^{2}\omega^{2} + A^{2}\Delta^{2} +
x(1-x)v^2\left(\mathbf{p}-\mathbf{q}\right)^{2}, \\
J_{3} &=& 2A^{2}\omega^{2} + A^{2}\Delta^{2} +
2x(1-x)v^2\left(\mathbf{p}-\mathbf{q}\right)^{2}.
\end{eqnarray}

First, we assume $g = 0$ and analyze the influence of disorder in
the normal phase of 3D DSM. The dependence of $A(\omega)$ and
$\omega A(\omega)$ on $\omega$ are displayed in
figures~\ref{Fig:AFun3DDirac}(a) and (b) by the dashed lines. We can find $A(\omega)$ is
saturated to a finite value if $\gamma$ is small. Whereas,
$A(\omega)$ becomes divergent and $\omega A(\omega)$ approaches to a
finite value provided $\gamma$ is large enough. These results are
qualitatively the same as those obtained by ignoring the vertex
correction presented in figures~\ref{Fig:AFun3DDirac}(a) and (b) by the solid lines.
However, the magnitude of $\gamma_{c}$ becomes smaller as the vertex
correction is taken into account, which can be easily observed from
figure~\ref{Fig:3DPhaseDiagram}. We notice that, a recent functional
RG analysis \cite{Sbierski17} incorporated the vertex correction and
argued that the critical value of disorder strength is smaller than
that obtained by using the SCBA. Ominato and Koshino
\cite{Ominato15} also emphasized the importance of the vertex
correction in the estimate of disorder scattering rate.

\begin{figure}
\center
\includegraphics[width=2.2in]{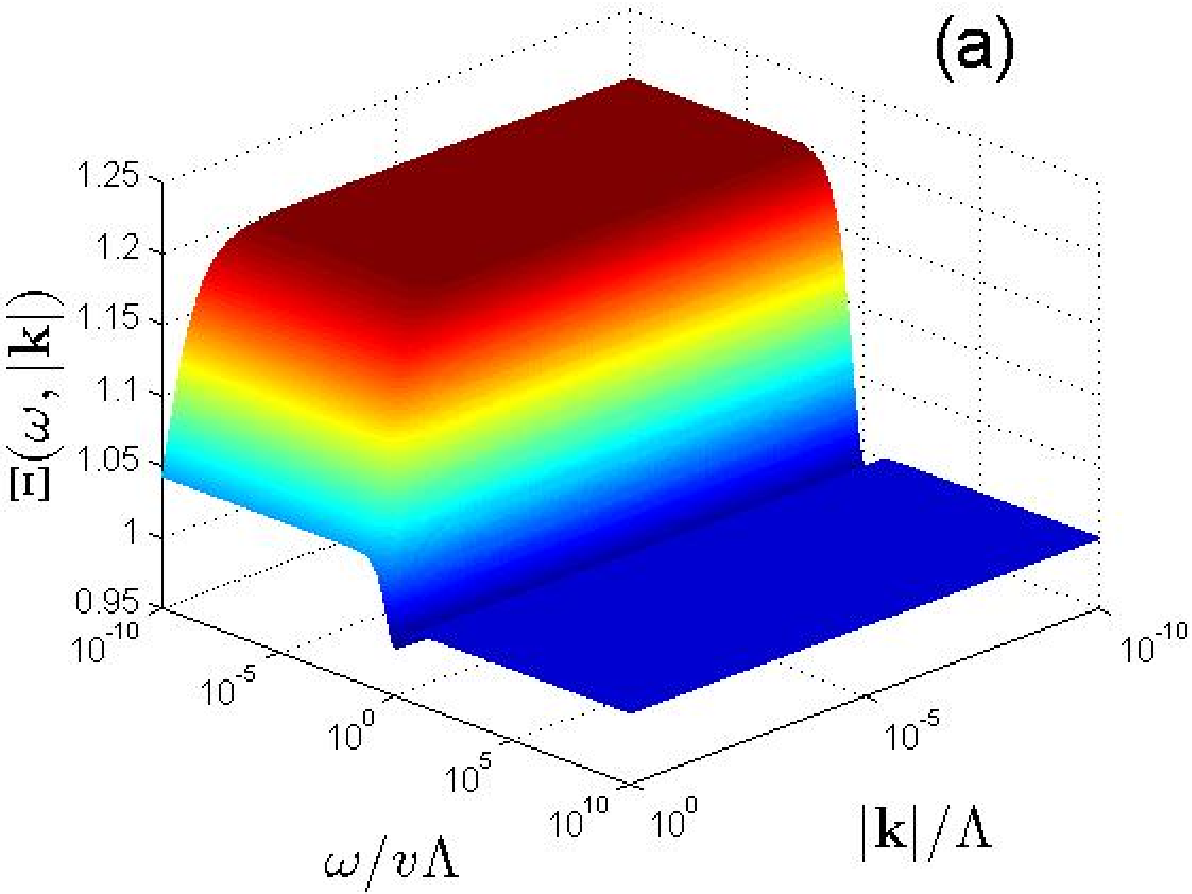}
\includegraphics[width=2.2in]{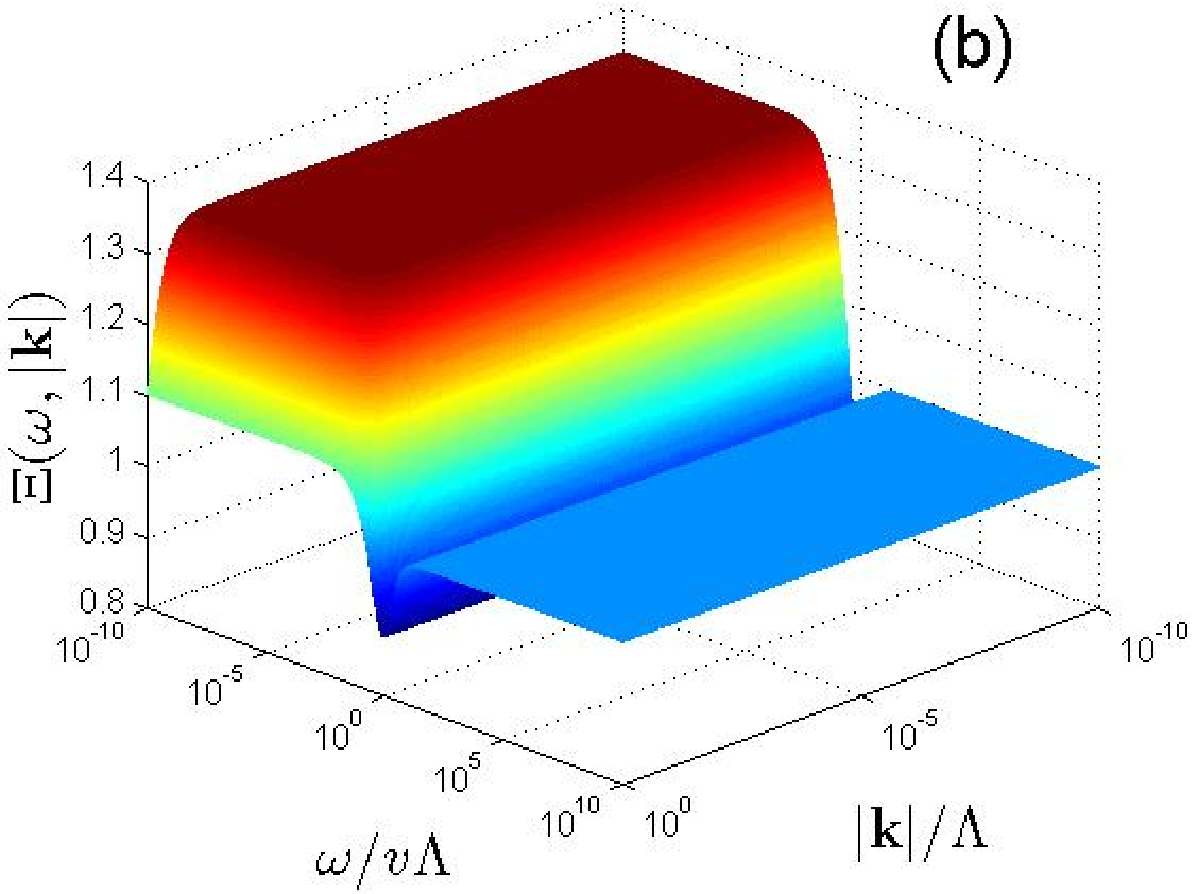}
\includegraphics[width=2.2in]{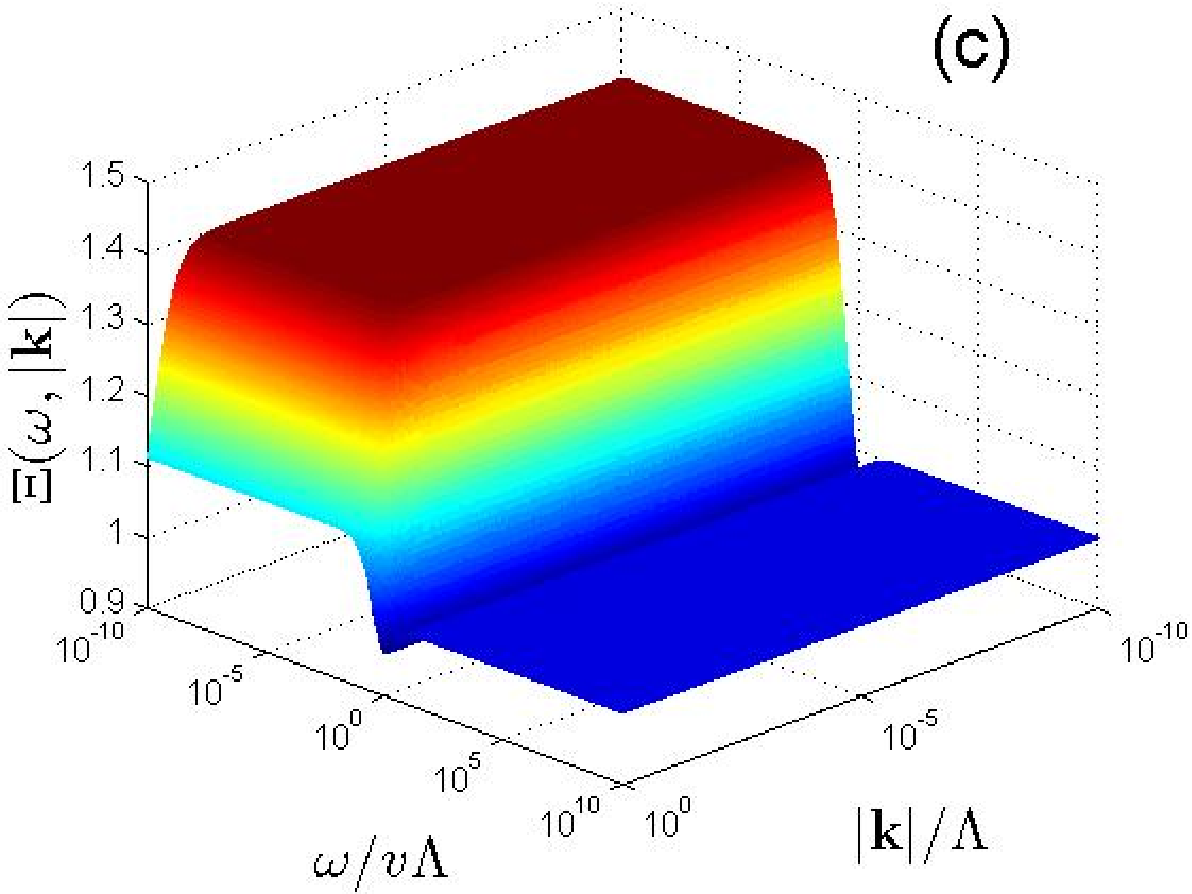}
\includegraphics[width=2.2in]{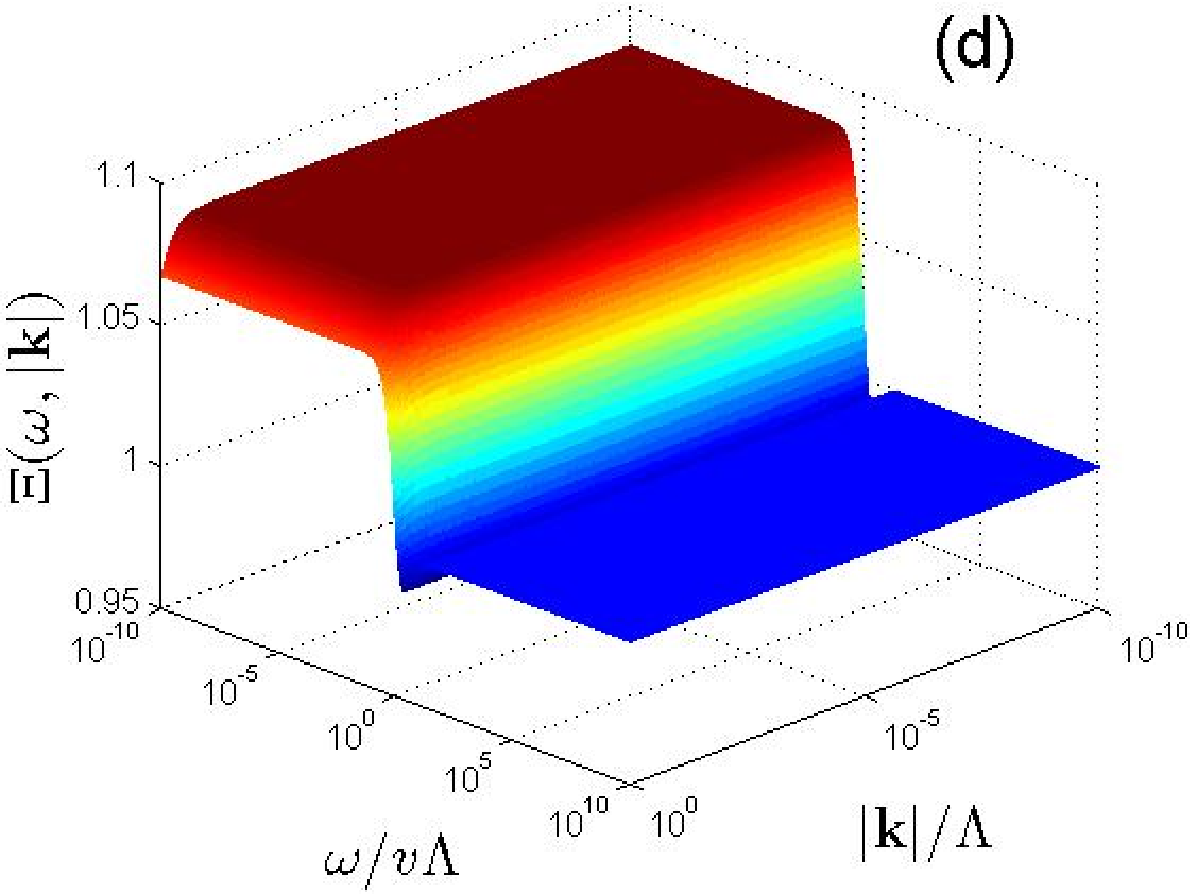}
\caption{Dependence of vertex function $\Xi(\omega,|\mathbf{k}|)$ on
the energy and momentum with $\gamma=0.2$ and $\gamma=1.5$ in (a)
and (b) respectively for 3D DSM. In (a) and (b), $\Delta = 0$ is
assumed. Dependence of vertex correction $\Xi(\omega,|\mathbf{k}|)$
on the energy and momentum with $g=0.9$ and $g=2$ in (c) and (d)
respectively for 3D DSM. In (c) and (d), we choose $\gamma = 0.5$.}
\label{Fig:Vertex3DDirac}
\end{figure}

Through numerical calculations of the
equations~(\ref{Eq:AFun3DDiracVertex}),
(\ref{Eq:VertexCorrection3DDirac}), and (\ref{Eq:Gap3DDirac}), we
obtain the phase diagram shown in figure~\ref{Fig:3DPhaseDiagram}(d).
According to this phase diagram, $g_{c}$ becomes smaller as $\gamma$
increases. Comparing figure~\ref{Fig:3DPhaseDiagram}(b) to
figure~\ref{Fig:3DPhaseDiagram}(d), we observe that the area of the SC
phase is broadened when the vertex correction is incorporated.

The dependence of $\Delta$ on $g$ is given in
figures~\ref{Fig:Gap3D}(a) and (b) by the dashed lines, and the dependence of $\Delta$ on
$\gamma$ in figures~\ref{Fig:Gapgamma3D}(a) and (b) by the dashed lines. The gap $\Delta$
is enhanced by weak disorder and suppressed by sufficiently strong
disorder when $g$ is small, but is always suppressed by disorder if
$g$ takes a large value. These results are qualitatively the same as
those obtained by ignoring the vertex correction. From
figures~\ref{Fig:Gap3D}, we can find that
the enhancement effect of the SC gap induced by weak
disorder for small $g$ is even more significant when the vertex
correction is considered.

According to figure~\ref{Fig:AFunGapPhase3DDirac}, $A(\omega)$ is
still saturated to certain finite constant. This is qualitatively
the same as the case of neglecting the vertex correction.
Quantitatively, $A(\omega)$ does acquire certain amount of
modification after including the vertex correction.

The behavior of $\Xi(\omega,|\mathbf{k}|)$ is presented in
figure~\ref{Fig:Vertex3DDirac}. In figures~\ref{Fig:Vertex3DDirac}(a)
and (b), we take $\Delta = 0$ by assuming that $g = 0$. According to
figures~\ref{Fig:Vertex3DDirac}(a) and (b), $\Xi$ is amplified when
$\gamma$ becomes larger. From figures~\ref{Fig:Vertex3DDirac}(c) and
(d), we see that in the SC phase, $\Xi$ decreases with
growing $g$, indicating that the vertex correction becomes less
important once a finite gap is opened.

\section{Superconductivity in 2D semi-Dirac semimetal \label{Sec:2DSemiDSM}}

The dispersion for 2D semi-Dirac fermions is given by
\begin{eqnarray}
E=\pm\sqrt{a^{2}k_{x}^{4}+v^2k_{y}^{2}}.
\end{eqnarray}
This dispersion is in between the fermion dispersions of ordinary 2D
metal and 2D DSM, thus the corresponding materials is usually called
2D semi-DSM. Such a type of fermions might emerge at the QCP between
a 2D DSM and a band insulator upon merging two separate Dirac points
to a single one.

Generation of semi-Dirac fermions through merging pairs of Dirac
points is theoretically predicted to exist in deformed graphene
\cite{Hasegawa06, Dietl08, Goerbig08, Montambaux09B}, pressured
organic compound $\alpha$-(BEDT-TTF)$_{2}$I$_{3}$ \cite{Goerbig08,
Montambaux09B, Kobayashi11}, few-layer black phosphorus that is
subject to a perpendicular electric field \cite{Dolui15, Yuan16} or
doping \cite{Baik15}, and also some artificial optical lattices
\cite{Wunsch08, Lim12}. Experimentally, the merging of Dirac points
and the appearance of semi-Dirac fermions were observed in ultracold
Fermi gas of $^{\mathrm{40}}$K atoms arranged on a honeycomb lattice
\cite{Tarruell12} and microwave cavities with graphene-like
structure \cite{Bellec13}. Kim \emph{et al.} \cite{Kim15} realized
semi-DSMs in few-layer black phosphorus at critical surface doping
with potassium. Moreover, robust semi-DSM state was predicted to
emerge in TiO$_{2}$/VO$_{2}$ nanostructure under suitable conditions
\cite{Pardo09, Banerjee09}. It was suggested by first-principle
calculations that semi-Dirac fermions are the low-energy excitations
of the strained puckered arsenene \cite{Kamal15, CanWang16}. In
addition, semi-DSM state may also be realized at the QCP between
normal insulator and topological insulator, and the QCP between
normal insulator and 2D DSM in time-reversal invariant 2D
noncentrosymmetric system \cite{AhnYang17}.

Recently, Uchoa and Seo \cite{Uchoa17} studied the possibility of
Cooper pairing in 2D semi-DSM by making a mean field analysis, and
argued that $s$-wave superconductivity is favored. Close to the QCP
between SM and SC phases, they found that the anisotropy of
quasiparticles leads to a novel smectic state of SC stripes. Roy and
Foster \cite{Roy17} investigated the influence of various
short-range interactions on the low-energy behavior of 2D semi-DSM
by making a RG analysis, and also discovered an $s$-wave
superconductivity. In the non-interacting limit, recent SCBA
\cite{Carpentier13} and RG \cite{Carpentier13, ZhaoPengLu16} studies
showed that arbitrarily weak random chemical potential turns the 2D
semi-DSM into a CDM state, analogous to what happens in 2D DSM.

Like 2D and 3D DSMs, the 2D semi-DSM also has a vanishing
zero-energy DOS if the chemical potential is tuned exactly at the
touching points. Consequently, the original AG approximation is no
longer valid. In what follows, we will examine how weak random
chemical potential affects the $s$-wave superconductivity in a 2D
semi-DSM by means of the AG approach and its generalization. Once
again, we will not give the mean-field Hamiltonian, and start our
discussion directly from the gap equation.

\subsection{Clean Case}

In the clean limit, the equation for the SC gap takes the form
\begin{eqnarray}
\Delta = 2g\int\frac{d\omega}{2\pi}\int
\frac{d^2\mathbf{k}}{(2\pi)^{2}}\frac{\Delta}{\omega^{2} +
a^{2}k_{x}^{4}+v^{2}k_{y}^{2}+\Delta^{2}},
\end{eqnarray}
which can be further written as
\begin{eqnarray}
1 = g\frac{1}{\pi^{2}}\int d|k_{x}| d|k_{y}|
\frac{1}{\sqrt{a^{2}k_{x}^{4}+v^{2}k_{y}^{2}+\Delta^{2}}}.
\end{eqnarray}
We employ the transformations
\begin{eqnarray}
E = \sqrt{a^2k_{x}^{4} + v^{2}k_{y}^{2}}, \qquad \delta =
\frac{ak_{x}^{2}}{v\left|k_{y}\right|},\label{Eq:IntegalTransformA}
\end{eqnarray}
which are equivalent to
\begin{eqnarray}
\left|k_{x}\right| = \frac{\sqrt{\delta}\sqrt{E}}{\sqrt{a} \left(1 +
\delta^2\right)^{\frac{1}{4}}}, \qquad \left|k_{y}\right| =
\frac{E}{v\sqrt{1+\delta^2}}.\label{Eq:IntegalTransformB}
\end{eqnarray}
The measures of integration satisfy the relation
\begin{eqnarray}
d|k_{x}|d|k_{y}|&=&
\left|\left|
\begin{array}{cc}
\frac{\partial |k_{x}|}{\partial E} & \frac{\partial
|k_{x}|}{\partial \delta} \\
\frac{\partial |k_{y}|}{\partial E} & \frac{\partial
|k_{y}|}{\partial \delta}
\end{array}
\right|\right|dEd\delta
= \frac{\sqrt{E}}{2v\sqrt{a}\sqrt{\delta}\left(1 +
\delta^2\right)^{3/4}}dEd\delta.
\label{Eq:IntegralTranformMeasure}
\end{eqnarray}
Adopting the transformations equations~(\ref{Eq:IntegalTransformB}) and
(\ref{Eq:IntegralTranformMeasure}), the gap equation becomes
\begin{eqnarray}
1 &=& g\frac{1}{2\pi^{2}v\sqrt{a}}
\int_{0}^{\Lambda_{E}}\frac{\sqrt{E}}{\sqrt{E^{2}+\Delta^{2}}} dE
\int_{0}^{+\infty}d\delta\frac{1}{\sqrt{\delta}\left(1 +
\delta^2\right)^{3/4}} \nonumber \\
&=&g\frac{\Gamma\left(\frac{5}{4}\right)}{\pi^{3/2}
\Gamma\left(\frac{3}{4}\right)v\sqrt{a}}\int_{0}^{\Lambda_{E}}
\frac{\sqrt{E}}{\sqrt{E^{2}+\Delta^{2}}} dE,
\end{eqnarray}
where $\Lambda_{E}$ is a cutoff for the variable $E$. Taking the
limit $\Delta \rightarrow 0$, we obtain the following critical value
\begin{eqnarray}
g_{c0} = \frac{\pi^{3/2}\Gamma\left(\frac{3}{4}\right)v
\sqrt{a}}{2\Gamma\left(\frac{5}{4}\right)\sqrt{\Lambda_{E}}},
\end{eqnarray}
beyond which a nonzero SC gap is opened. This $g_{c0}$
is the QCP that separates the SM and SC phases.

\begin{figure}
\center
\includegraphics[width=2.2in]{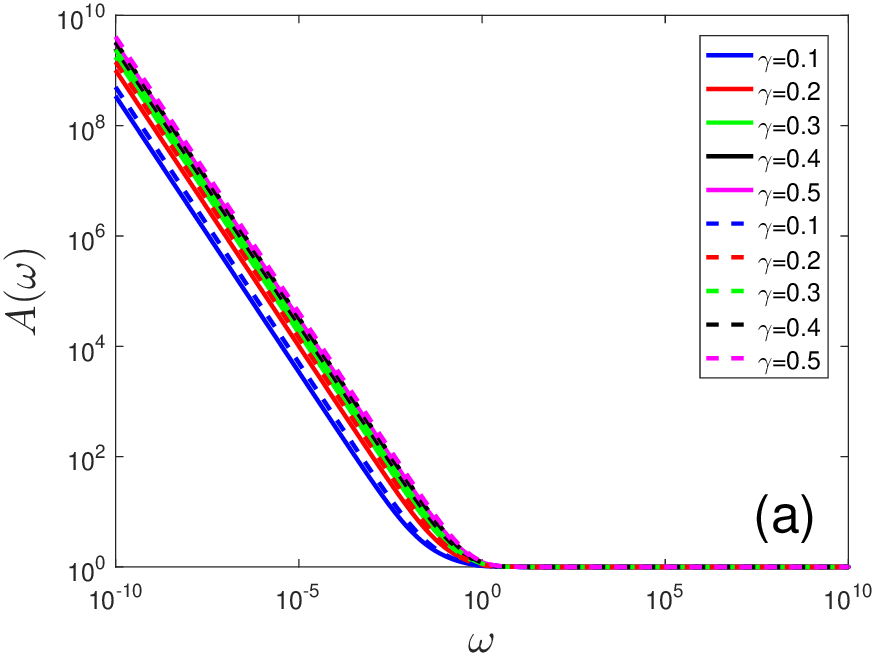}
\includegraphics[width=2.2in]{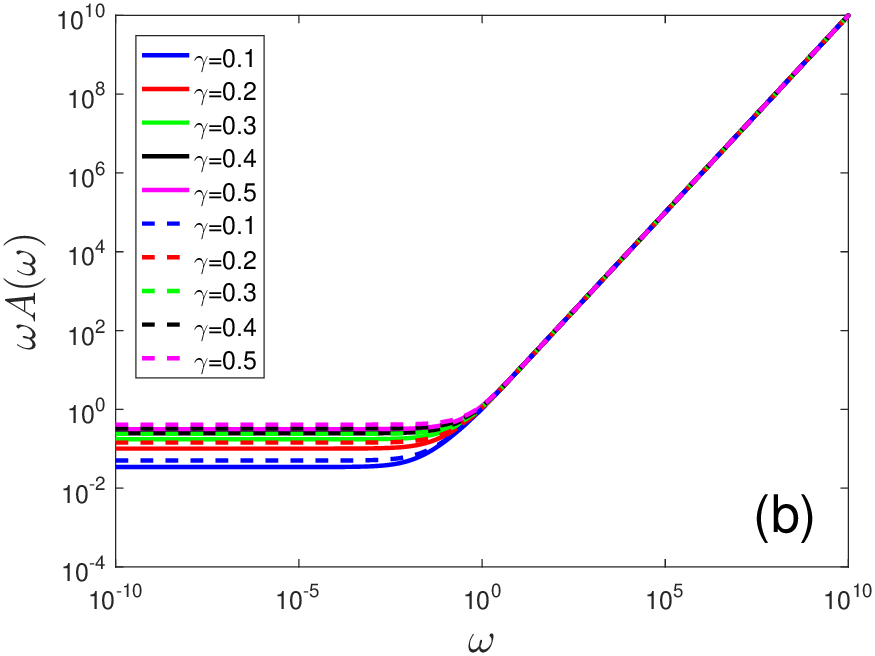}
\caption{(a) Dependence of $A$ and (b) $\omega A$ on $\omega$
with different values of $\gamma$ for 2D semi-DSM. Vertex correction is
neglected for solid lines, but incorporated for dashed lines. $\Delta =
0$ is taken. \label{Fig:AFunSemiDirac}}
\end{figure}

\subsection{Analysis by AG method without vertex correction}

Including the disorder scattering, the self-consistent equations
obtained under the original AG approximation are found to have the
forms
\begin{eqnarray}
A &=& 1+\gamma A F(\omega,\Delta), \label{Eq:AGEqSemiDSMA}
\\
1 &=&\frac{g}{g_{c0}}\frac{1}{2\pi}\int_{-\infty}^{+\infty} d\omega
A(\omega) F(\omega,\Delta), \label{Eq:AGEqSemiDSMB}
\end{eqnarray}
where
\begin{eqnarray}
&&\gamma =\frac{\Gamma\left(\frac{5}{4}\right)
n_{\mathrm{imp}}u^{2}}{\pi^{3/2}\Gamma\left(\frac{3}{4}\right)
v\sqrt{a}\sqrt{\Lambda_{E}}}, \\
&&F(\omega,\Delta) = \frac{1}{\sqrt{2}J_{4}^{1/4}}
\left[\frac{1}{2}\ln\left(\frac{J_{4}^{1/2}-\sqrt{2}J_{4}^{1/4} +
1}{J_{4}^{1/2}+\sqrt{2} J_{4}^{1/4}+1}\right)+ \arctan\left(1 +
\frac{\sqrt{2}}{J_{4}^{1/4}} \right)\right.\nonumber
\\
&&\left. - \arctan\left(1-\frac{\sqrt{2}}{J_{4}^{1/4}}
\right)\right],
\end{eqnarray}
with $J_{4} = A^{2}\omega^{2} + A^{2}\Delta^{2}$. The scaling
transformations $\frac{\omega}{\Lambda_{E}}\rightarrow \omega$ and
$\frac{\Delta}{\Lambda_{E}}\rightarrow\Delta$ have been used.

\begin{figure}
\center
\includegraphics[width=2.2in]{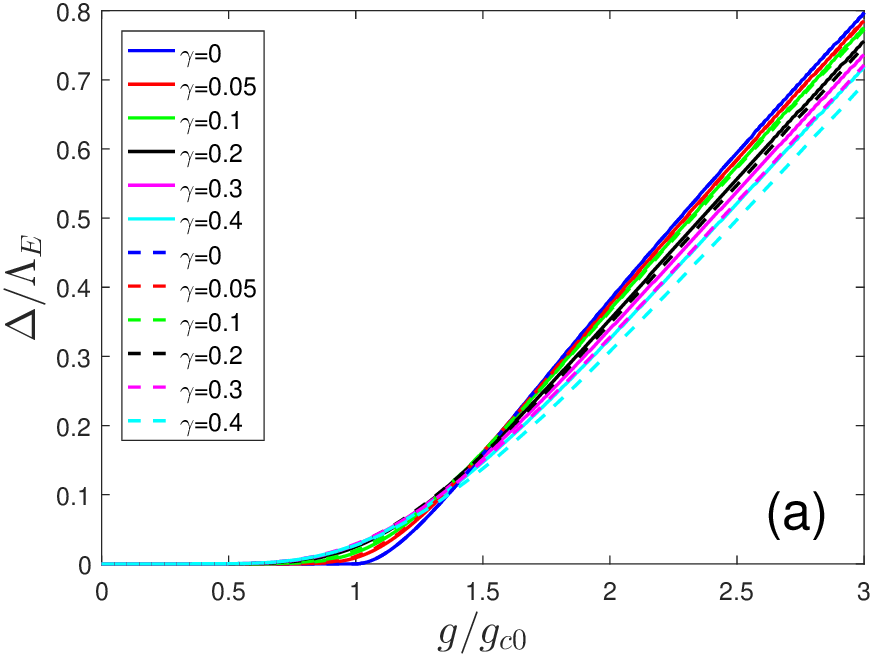}
\includegraphics[width=2.2in]{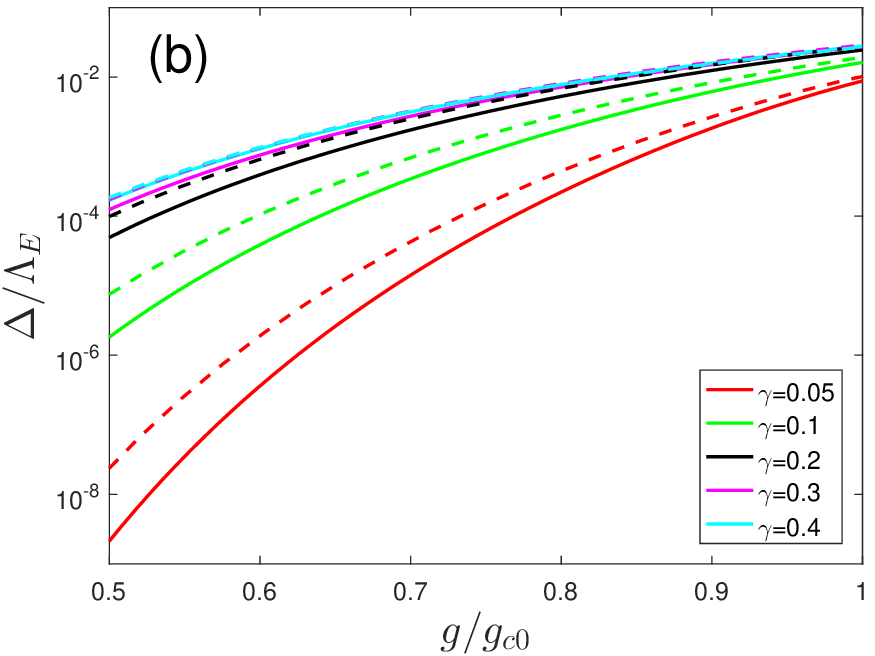}
\caption{Dependence of $\Delta$ on $g$ at different values of
impurity strength $\gamma$ for 2D semi-DSM. Vertex correction is
neglected for solid lines, but incorporated for dashed lines.
\label{Fig:GapSemiDirac}}
\end{figure}

By taking $\Delta = 0$, we obtain the solution of $A(\omega)$ in the
normal state, and show the results in
figures~\ref{Fig:AFunSemiDirac}(a) and (b) by the solid lines. We find that $A(\omega)$
is divergent in the lowest energy limit, and $\omega A(\omega)$
approaches to a constant in 2D semi-DSM, similar to 2D DSM. In the
limit $\Delta=0$, the integrand in the equation~(\ref{Eq:AGEqSemiDSMB})
satisfies
\begin{eqnarray}
A(\omega)F(\omega,0)\propto\frac{\Gamma}{|\omega|}
\end{eqnarray}
in the low energy regime. It is easy to see that the integration in
the equation~(\ref{Eq:AGEqSemiDSMB}) is divergent, which in turn means
$g_{c}/g_{c0}\rightarrow 0$. Therefore, superconductivity is
produced by arbitrarily weak pairing interaction once random
chemical potential is considered. Dependence of $\Delta$ on $g$ with
different values of $\gamma$ is presented in
figures~\ref{Fig:GapSemiDirac}(a) and (b) by the solid lines, which clearly exhibits the
promotion of superconductivity due to weak random chemical potential
if $g$ is small.

In the SC phase with $\Delta \neq 0$, the dependence of $\Delta$ on
$\gamma$ is depicted in figures~\ref{Fig:GapgammaSemiDirac}(a) and
(b) by the solid lines. For small values of $g$, the gap $\Delta$ increases as $\gamma$
is growing, and then starts to decrease with growing $\gamma$ when
$\gamma$ becomes sufficiently large. For larger values of $g$,
$\Delta$ decreases monotonously as $\gamma$ increases. We thus can
see that the influence of random chemical potential on $s$-wave
superconductivity in 2D semi-DSM is qualitatively the same as 2D
DSM.

\subsection{Analysis beyond AG approximation}

In this subsection, we move to examine the impact of the vertex
correction. After including the vertex correction, the equation for
$A$ is of the form
\begin{eqnarray}
A &=& 1 + \gamma \frac{\Gamma\left(\frac{3}{4}\right)}{\pi^{
1/2}\Gamma\left(\frac{5}{4}\right)}A\int_{0}^{1} dk_{x}
\int_{0}^{1}dk_{y}\frac{\Xi\left(\omega,|k_{x}|,|k_{y}|\right)}{A^{2}
\omega^{2} + k_{x}^{4}+k_{y}^{2}+A^{2}\Delta^{2}},
\end{eqnarray}
where the vertex function $\Xi$ is given by
\begin{eqnarray}
\fl &&\Xi\left(\omega,\left|p_{x}-q_{x}\right|,|p_{y}-q_{y}|\right)
= 1+\gamma \frac{\pi^{1/2} \Gamma\left(\frac{3}{4}
\right)}{8\Gamma\left(\frac{5}{4}\right)} \int_{0}^{1}dx\nonumber
\\
\fl &&\times\left[\int_{0}^{1} dk_{x} \frac{\left(k_{x} +
\left|p_{x}-q_{x}\right|\right)^{2}k_{x}^{2} + x\left(k_{x} +
\left|p_{x}-q_{x}\right|\right)^{4}+(1-x)k_{x}^{4} +
2A^{2}\Delta^{2}}{\left[A^{2}\omega^{2} + x\left(k_{x} +
\left|p_{x}-q_{x}\right|\right)^{4}+(1-x)k_{x}^{4} +
A^{2}\Delta^{2}+x(1-x)\left|p_{y} -
q_{y}\right|^{2}\right]^{3/2}}\right.\nonumber
\\
\fl &&\left.+\int_{0}^{1} dk_{x} \frac{\left(k_{x}-\left|p_{x} -
q_{x}\right|\right)^{2}k_{x}^{2} + x\left(k_{x}-\left|p_{x} -
q_{x}\right|\right)^{4}+(1-x)k_{x}^{4}+2A^{2}\Delta^{2}}{\left[A^{2}
\omega^{2} + x\left(k_{x}-\left|p_{x}-q_{x}\right|\right)^{4} +
(1-x)k_{x}^{4} + A^{2}\Delta^{2}+x(1-x)\left|p_{y} -
q_{y}\right|^{2}\right]^{3/2}}\right].
\end{eqnarray}
We have used the transformations
\begin{eqnarray}
\fl \frac{k_{x}}{\Lambda_{x}}\rightarrow k_{x},\quad
\frac{k_{y}}{\Lambda_{y}}\rightarrow k_{y},\quad
\frac{\omega}{\Lambda_{E}}\rightarrow \omega,\quad
\frac{\Delta}{\Lambda_{E}}\rightarrow \Delta, \quad
\frac{v^2\left(p_{y}-q_{y}\right)^{2}}{\Lambda_{E}^{2}}\rightarrow
\left(p_{y}-q_{y}\right)^{2},
\end{eqnarray}
and also made the assumption that $a\Lambda_{x}^{2} = v\Lambda_{y} =
\Lambda_{E}$. The equation for the gap $\Delta$ is still given by
equation~(\ref{Eq:AGEqSemiDSMB}).

\begin{figure}
\center
\includegraphics[width=2.6in]{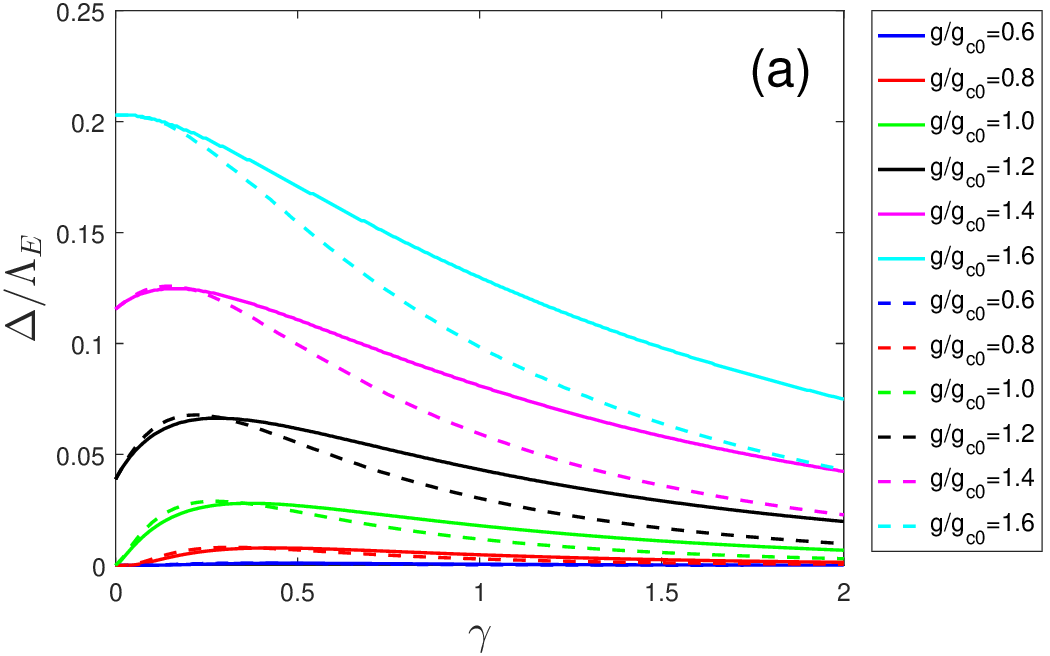}
\includegraphics[width=2.6in]{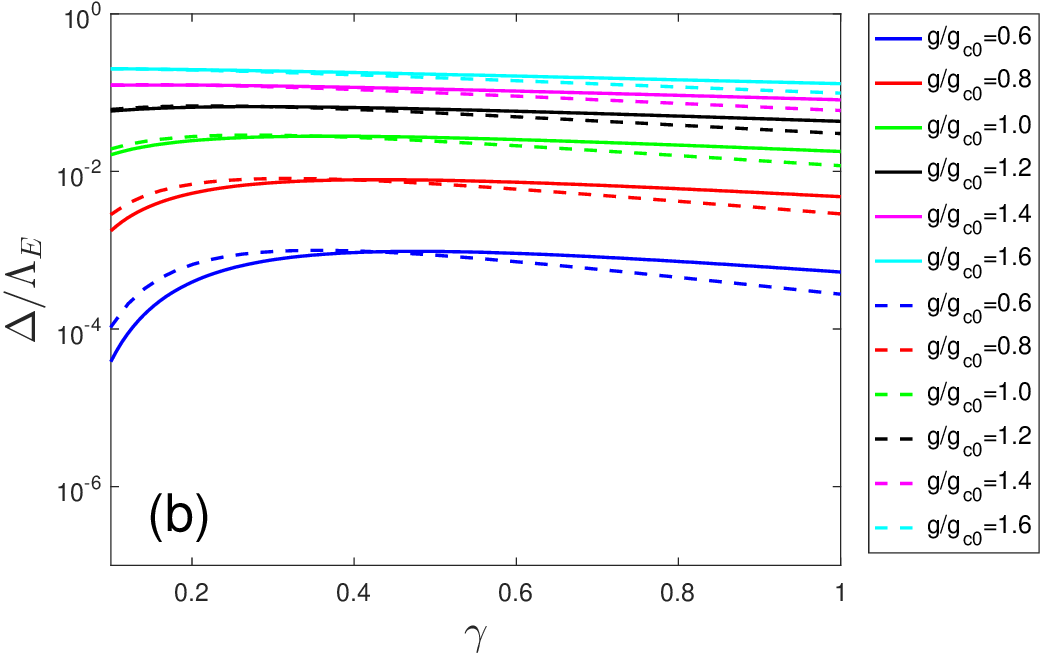}
\caption{Dependence of $\Delta$ on $\gamma$ at different values of
$g$ for 2D semi-DSM. Vertex correction is
neglected for solid lines, but incorporated for dashed lines. \label{Fig:GapgammaSemiDirac}}
\end{figure}

After including the vertex correction, we solve the self-consistent
equations, and show the dependence of $\Delta$ on $g$ at different
values of $\gamma$ in figures~\ref{Fig:GapSemiDirac}(a) and (b) by the dashed lines. The
relation between $\Delta$ and $\gamma$ at different values of $g$ is
displayed in figures~\ref{Fig:GapgammaSemiDirac}(a) and (b) by the dashed lines. We
observe from figure~(\ref{Fig:GapSemiDirac}) and
figure~(\ref{Fig:GapgammaSemiDirac}) that including the vertex
correction does not change the qualitative property of the influence
of random chemical potential on superconductivity in 2D semi-DSM,
but it leads to obvious quantitative increase of the SC gap in presence of weak attraction
and weak disorder. All these
results are similar to that obtained in the case of 2D DSM.

\section{Superconductivity in Bilayer graphene \label{Sec:BilayerGraphene}}

In bilayer graphene with Bernal AB stacking, in which two layers of
carbon atoms are rotated by 60$^{\circ}$, the Fermi surface is also
composed of discrete points \cite{CastoNeto09, Kotov12}. There are
also other sorts of configuration of bilayer graphene, such as AA
stacking that match the A sublattices of two layers. Here, we focus
on bilayer graphene with Bernal configuration, which is the most
frequently studied case. In such a system, the fermions display the
following dispersion \cite{CastoNeto09, Kotov12}
\begin{eqnarray}
E = \pm a|\mathbf{k}|^{2},
\end{eqnarray}
where $a$ is a constant. For Bernal bilayer graphene, the Berry
phase around the touching points is trivial, in distinct to
monolayer graphene that exhibits a non-trivial Berry phase. This
difference can be clearly observed in quantum Hall effect
experiments \cite{Novoselov06BLGraphene}. Due to its special
dispersion and dimensionality, the bilayer graphene has a finite
zero-energy DOS \cite{Kotov12}: $\rho(0)\propto 1/a$. As a result,
an infinitesimal short-range interaction is able to drive some type
of phase-transition instability \cite{Kotov12, Vafek10, ZhangFan10}.

\subsection{Clean Case}

In the clean limit, the gap equation for $s$-wave superconductivity
in bilayer graphene is given by
\begin{eqnarray}
\Delta = 2g\int\frac{d\omega}{2\pi}\int
\frac{d^2\mathbf{k}}{(2\pi)^{2}}\frac{\Delta}{\omega^{2} +
a^{2}k^{4}+\Delta^{2}},
\end{eqnarray}
where $g$ is the strength parameter of pairing interaction.
Performing the integrations of energy and momenta, we obtain
\begin{eqnarray}
1 = \frac{g}{4\pi a} \ln\left(\frac{\sqrt{a^{2}\Lambda^4 +
\Delta^{2}}+a\Lambda^2}{\Delta}\right),
\end{eqnarray}
where $\Lambda$ is the cutoff of the momentum. In the limit
$\Delta\rightarrow 0$, $g$ approaches a critical value $g_{c0}$ that
satisfies
\begin{eqnarray}
1 = \lim_{\Delta\rightarrow0}\frac{g_{c0}}{4\pi a}
\ln\left(\frac{2a\Lambda^2}{\Delta}\right).
\end{eqnarray}
It is obvious that $g_{c0}\rightarrow 0$, so arbitrarily weak
attraction leads to superconductivity in bilayer graphene. This
behavior is markedly different from the aforementioned SMs
with vanishing DOS at the Fermi level.

\subsection{Analysis by AG method without vertex correction}

In the absence of vertex correction, the self-consistent equations
for the renormalization factor $A(\omega)$ and the SC
gap $\Delta$ can be written as
\begin{eqnarray}
A &=& 1 + \gamma\frac{a\Lambda^{2}}{\sqrt{\omega^{2} +
\Delta^{2}}}\arctan\left(\frac{a\Lambda^{2}}{\sqrt{J_{4}}}\right),
\label{Eq:AGEqBLGrapheneA}
\\
1 &=&\frac{g}{g_{0}}\int d\omega \frac{1}{\sqrt{\omega^{2} +
\Delta^{2}}} \arctan\left(\frac{a\Lambda^{2}}{\sqrt{J_{4}}}\right),
\label{Eq:AGEqBLGrapheneB}
\end{eqnarray}
where
\begin{eqnarray}
\gamma = \frac{n_{\mathrm{imp}}u^{2}}{4\pi a^{2}\Lambda^{2}}, \qquad
g_{0} = 4\pi^{2} a,
\end{eqnarray}
and $J_{4} = A^{2}\omega^{2} + A^{2}\Delta^{2}$. Here, $g_{0}$ is
chosen as the unit of attraction strength $g$.

\begin{figure}
\center
\includegraphics[width=2.2in]{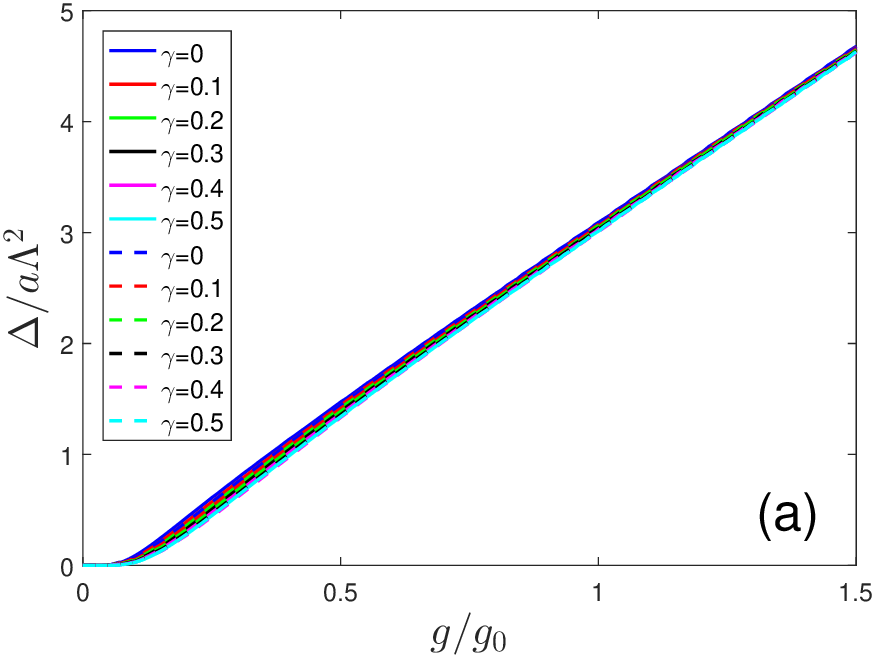}
\includegraphics[width=2.2in]{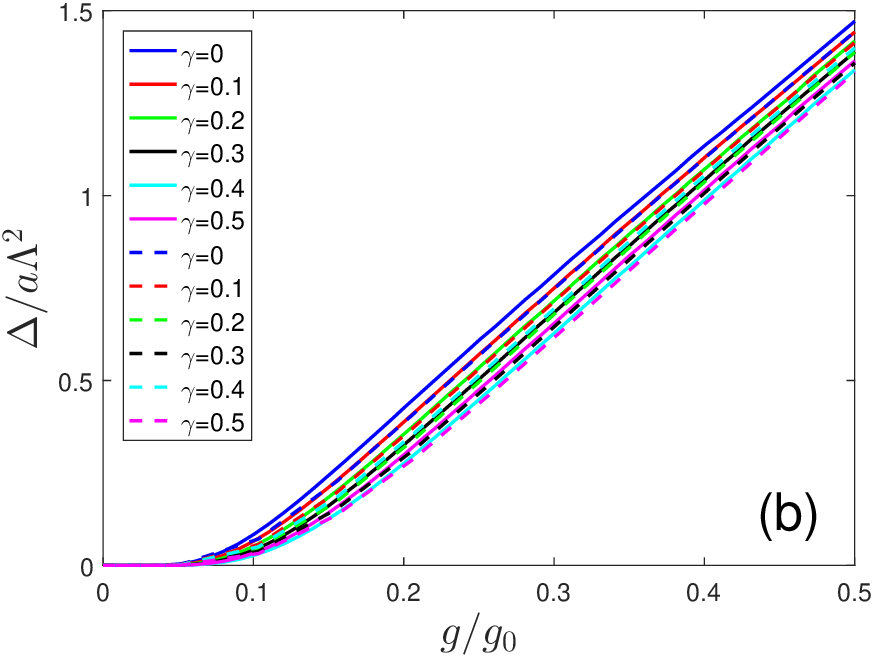}
\caption{Dependence of $\Delta$ on $g$ at different values of
impurity strength $\gamma$ for bilayer graphene. Vertex correction is
neglected for solid lines, but incorporated for dashed lines.
\label{Fig:GapBLGraphene}}
\end{figure}

\begin{figure}
\center
\includegraphics[width=2.6in]{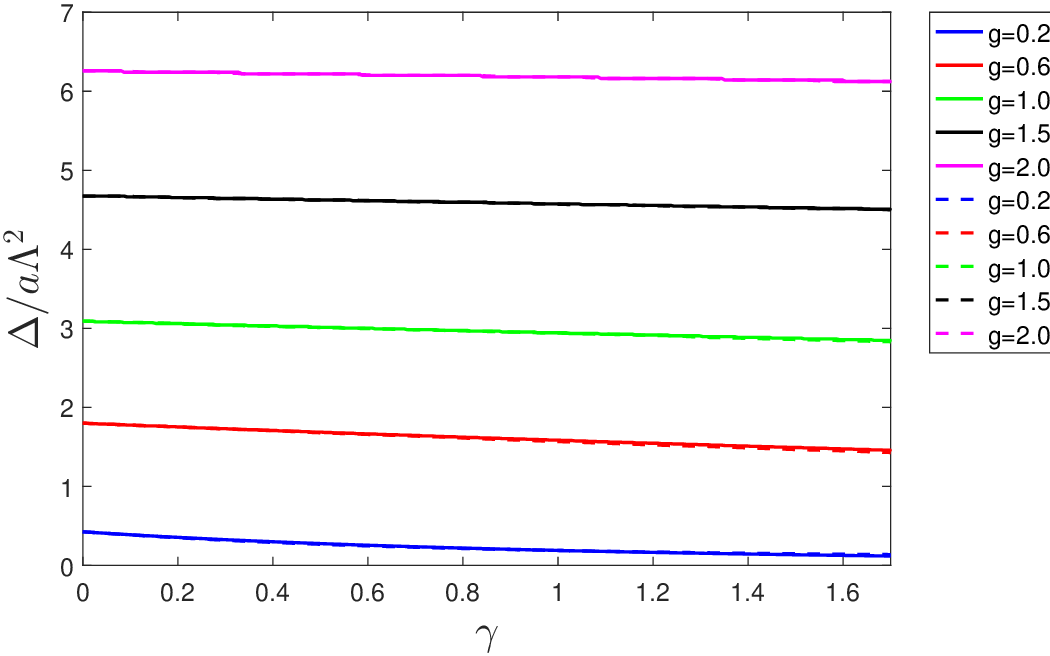}
\caption{Dependence of $\Delta$ on $\gamma$ at different values of
$g$ for bilayer graphene. Vertex correction is
neglected for solid lines, but incorporated for dashed lines.\label{Fig:GapBLGraphenegamma}}
\end{figure}

Solving equations~(\ref{Eq:AGEqBLGrapheneA}) and
(\ref{Eq:AGEqBLGrapheneB}), we obtain the dependence of $\Delta$ on
$g$ at different values of $\gamma$, which is shown in
figures~\ref{Fig:GapBLGraphene}(a) and (b) by the solid lines. The SC gap $\Delta$ is
quantitatively suppressed by random chemical potential. The relation
between $\Delta$ and $\gamma$ at different values of $g$ is
displayed in figure~\ref{Fig:GapBLGraphenegamma} by the solid lines. The gap $\Delta$
is suppressed monotonously by increasing $\gamma$, but the
suppression effect is not significant. Thus, the $s$-wave
superconductivity seems to be robust against weak disorder in
bilayer graphene. For bilayer graphene, there is not any evidence of
disorder-induced promotion of superconductivity, which occurs in 2D
DSM, 3D DSM, and 2D semi-DSM. Such difference originates from the
fact that the zero-energy DOS $\rho(0)$ is nonzero in bilayer
graphene, but vanishes in the other three types of SM.

\subsection{Analysis beyond AG approximation}

After including the vertex correction, we re-write the
self-consistent equations for the function $A$ and the vertex $\Xi$
as follows
\begin{eqnarray}
\fl && A = 1+2\gamma A\int_{0}^{1}dk k \frac{\Xi
\left(\omega,|\mathbf{k}|\right)}{A^{2}\omega^{2} + k^{4} +
A^{2}\Delta^{2}},
\\
\fl \Xi\left(\omega,\left|\mathbf{p}-\mathbf{q}\right|\right) &=&
1+\gamma\frac{1}{\pi}\int_{0}^{1}dk k\int_{0}^{2\pi}d\varphi
\nonumber \\
&&\times\left[-A^{2}\omega^{2}+\left(k^{2}+2k\left|\mathbf{p} -
\mathbf{q}\right|\cos \varphi + \left|\mathbf{p} -
\mathbf{q}\right|^{2}\right)k^{2}+A^{2}\Delta^{2}\right]\nonumber
\\
\fl &&\times\left[A^{2}\omega^{2}+\left(k^{4} +
4k^3\left|\mathbf{p}-\mathbf{q}\right|\cos \varphi +
4k^{2}\left|\mathbf{p}-\mathbf{q}\right|^{2}\cos^{2}\varphi +
2k^{2}\left|\mathbf{p}-\mathbf{q}\right|^{2} \right.\right.\nonumber
\\
\fl &&\left.\left.+4k\left|\mathbf{p}-\mathbf{q}\right|^{3}\cos \varphi
+ \left|\mathbf{p}-\mathbf{q}\right|^{4}\right) +
A^{2}\Delta^{2}\right]^{-1}\left[A^{2}\omega^{2} +
k^{4}+A^{2}\Delta^{2}\right]^{-1}.
\end{eqnarray}
The gap still satisfies the equation~(\ref{Eq:AGEqBLGrapheneB}). Including
the vertex correction, the relation between $\Delta$ and $g$ for
different values of $\gamma$ are shown in
figures~\ref{Fig:GapBLGraphene}(a) and (b) by the dashed lines. The dependence of $\Delta$
on $\gamma$ for different values of $g$ is depicted in
figure~\ref{Fig:GapBLGraphenegamma} by the dashed lines. We can find that vertex
correction does not lead to qualitative change of the results.
Quantitatively, the suppression of gap by disorder becomes only
slightly stronger once the vertex correction is considered, which
indicates that the vertex correction can be nearly neglected.

\section{Remarks on related issues \label{Sec:Discussions}}

In this section, we discuss several related issues.

\subsection{Truncation of DSEs}

As analyzed in the Introduction, disorder scattering and Cooper
pairing can affect each other substantially, and thus should be
treated self-consistently. The non-perturbative DSEs approach
provides an ideal formalism for such a self-consistent treatment. In
the past several decades, this approach has been widely applied to
investigate the non-perturbative phenomenon of dynamical symmetry
breaking in high-energy physics \cite{Roberts94, Roberts00}, the
dynamical chiral symmetry breaking in three-dimensional quantum
electrodynamic (QED$_{3}$) \cite{Appelquist88, Nash89, Liu03,
Fisher04} which is effective model in several important condensed
matter systems, the excitonic insulating transition in various SM
materials \cite{Khveshchenko01, Gorbar02, Liu09, Gamayun10,
WangLiu12, Popovici13, Gonzalez15, Carrington16, Gonzalez14,
Gonzalez17, Janssen15, WangLiuZhang17}. Moreover, the DSEs approach
has been applied to study the interplay of non-Fermi liquid behavior
and SC pairing in various strongly correlated systems
\cite{Abanov01, WangZhiQiang01, AbanovChubukov03, Chubukov05,
Yamase13, Lederer15, Einenkel15, WangYuXuan16, WangHuaJia17A,
WangXiaoYu17, LiuWang17, Chung13, WangChakravarty16, Isobe17}. The
most noticeable advantage of DSEs approach is that it provides a
non-perturbative framework to quantitatively calculate various
physical quantities, such as the Landau damping rate, disorder
scattering rate, and SC gap \emph{etc.}, by incorporating several
types of interactions in a self-consistent manner.

In the Ref.~\cite{ZhangJunHua15}, Zhang \emph{et al.} studied the
impact of nonmagnetic short-range disorder on some ordered states by
solving a special set of self-consistent equations, which is similar
to the AG method. Their main finding is that, the fully gapped state
is suppressed by disorder more significantly than the nematic state,
and disorder may induce a quantum phase transition between a fully
gapped ordered state and a nematic state. Such results might account
for the discrepancy between experiments of bilayer graphene. The
DSEs method was also applied  to investigate the interplay of
disorder scattering and Coulomb interaction in 3D DSM \cite{Gonzalez17}.

In the application of DSEs method to the superconductivity in
disordered systems, one needs to construct and solve a set of
non-linear integral equations for the gap $\Delta$ and disorder
scattering rate $\Gamma$. In its most generic form, the DSEs are
exact and contain all the physical processes. However, solving the
complete set of DSEs is impossible. In practice, it is always
necessary to employ certain truncation, which is implemented by
retaining the most important contribution and discarding some higher
order contributions. All the existing DSEs studies \cite{Roberts94,
Roberts00, Appelquist88, Nash89, Liu03, Fisher04, Khveshchenko01,
Gorbar02, Liu09, Gamayun10, WangLiu12, Popovici13, Gonzalez15,
Carrington16, Gonzalez14, Gonzalez17, Janssen15, WangLiuZhang17,
Abanov01, WangZhiQiang01, AbanovChubukov03, Chubukov05, Yamase13,
Lederer15, Einenkel15, WangYuXuan16, WangHuaJia17A, WangXiaoYu17,
LiuWang17, Chung13, WangChakravarty16, Isobe17} adopt the following
strategy: consider the lowest order truncation to capture the key
physical picture; include the higher order corrections step by step
to examine the validity of the conclusion obtained by the lowest
order calculation. The AG method \cite{AbrikosovGorkov59} can be
regarded as the lowest order truncation of DSEs of disorder
scattering and Cooper pairing. Under the original AG approximation,
the fermion self-energy is calculated at the leading order, with all
vertex corrections entirely ignored. We emphasize here that,
although the AG method neglects most higher order corrections, it is
non-perturbative in nature and can treat the mutual influence
between disorder scattering and Cooper pairing equally. When the AG
method is applied to disordered SM materials, the vertex correction
may no longer be simply ignored. In this paper, we find that,
including vertex correction does not alter the qualitative
conclusion obtained by the original AG method, but leads to
considerable enhancement of SC gap size in 2D DSM, 3D DSM, and 2D
semi-DSM in weak-attraction regime.

To examine the validity of our conclusion, one might endeavor to
include even higher order corrections into the self-consistent
equations studied in this paper. This work is interesting, but out
of the scope of the present paper. The main difficulty is that, the
self-consistent DSEs become very complicated and are hard to solve
numerically. Here we only make a brief remark on the possible
influence of such corrections. From the results presented in the
last several sections, we can infer that the vertex correction
becomes progressively less important as the SC gap grows. For large
SC gap, the vertex correction can be safely ignored. Actually, the
SC gap provides an infrared cutoff, which regularizes the infrared
behavior and as such suppresses higher order corrections. When the
SC gap is not large, the higher order corrections omitted in our
analysis might more or less modify our results quantitatively. We
leave this project for future research.

Apart from including higher order corrections by brute force, our
conclusion, which states that random chemical potential promotes
superconductivity in 2D DSM, 3D DSM, and 2D semi-DSM, could also be
verified by experiments. The SC gap can be detected in scanning
tunneling microscopy (STM) measurements. One might prepare a series
of different SM materials to test whether it is easier to
achieve superconductivity in more disordered materials. We expect
that future experiments would be performed to clarify the
reliability of our conclusion.

\subsection{Contribution of two special diagrams}

\begin{figure}[htbp]
\center
\includegraphics[width=3.2in]{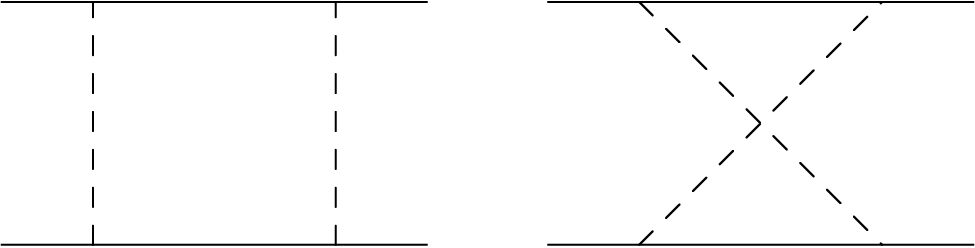}
\caption{Two possible diagrams for the correction to
fermion-disorder coupling. \label{Fig:TwoDiagrams}}
\end{figure}

One might think that the two Feynaman diagrams, shown in
figure~\ref{Fig:TwoDiagrams}, should be taken into account in the
calculation of the corrections to fermion-disorder coupling. We now
explain why these two diagrams are neglected.

In our work, we consider only one type of disorder, namely random
chemical potential. For 2D DSM and 3D DSM that contain only one type
of disorder, the contributions from these two Feynman diagrams
cancel, which is discussed in Ref.~\cite{Roy14}. For 2D semi-DSM and
bilayer graphene, these two diagrams can dynamically generate other
types of disorder, such as random gauge potential and random mass
\cite{Carpentier13}. To simplify the problem, we truncate the DSEs
by neglecting the dynamically generated disorder. An important point
is that random chemical potential always dominates the dynamically
generated random gauge potential and random mass. Including the
dynamically generated disorders would further enhance random
chemical potential and then induce a larger DOS $\rho(0)$ in the
normal state. For 2D semi-DSM, according to our results, a larger
disorder-induced DOS $\rho(0)$ would lead to stronger enhancement
superconductivity in presence of weak pairing interaction. In the
case of bilayer graphene that has a finite $\rho(0)$ even in the
clean limit, increasing the strength of random chemical potential
also does not change our basic conclusion that superconductivity is
slightly suppressed.

\subsection{Rare region effect}

In this article, we calculate the magnitude of SC gap, which can be
measured by STM experiments \cite{Lang02, Fischer07}. We do not
consider the rare region effect here. In Ref.~\cite{Nandkishore13},
Nandkishore \emph{et al.} studied the enhancement of
superconductivity by the rare region effect of disorder in 2D DSM
\cite{Nandkishore13}. They showed that local superconductivity is
substantially enhanced in the regions with stronger disorder and
larger local DOS. In case the Josephoson coupling between locally SC
regions is strong enough to establish global phase coherence, the
system becomes SC globally. After examining the rare region effect,
they obtained an obviously larger value of $T_{c}$, comparing to the
one calculated by using the mean-field analysis (AG method).

Nandkishore \emph{et al.} later studied the rare region effect in 3D
DSM \cite{Nandkishore14}, and concluded that arbitrarily weak random
chemical potential can induce a finite $\rho(0)$ due to rare region
effect. They also considered the interplay of random chemical
potential and superconductivity in 3D DSM. Superconductivity is
triggered only when the strength of attraction exceeds a threshold
in the clean system. However, local pairing can be triggered in rare
regions where the local DOS is nonzero, which drives the system into
a SC state once phase coherence between the islands is established
by Josephson coupling. These results suggest that superconductivity
is promoted by random chemical potential in 3D DSM, which is
qualitatively consistent with our conclusion presented in
section~\ref{Sec:3DDSM}.

\subsection{The ratio $\Delta/T_{c}$}

\begin{figure}
\center
\includegraphics[width=2.2in]{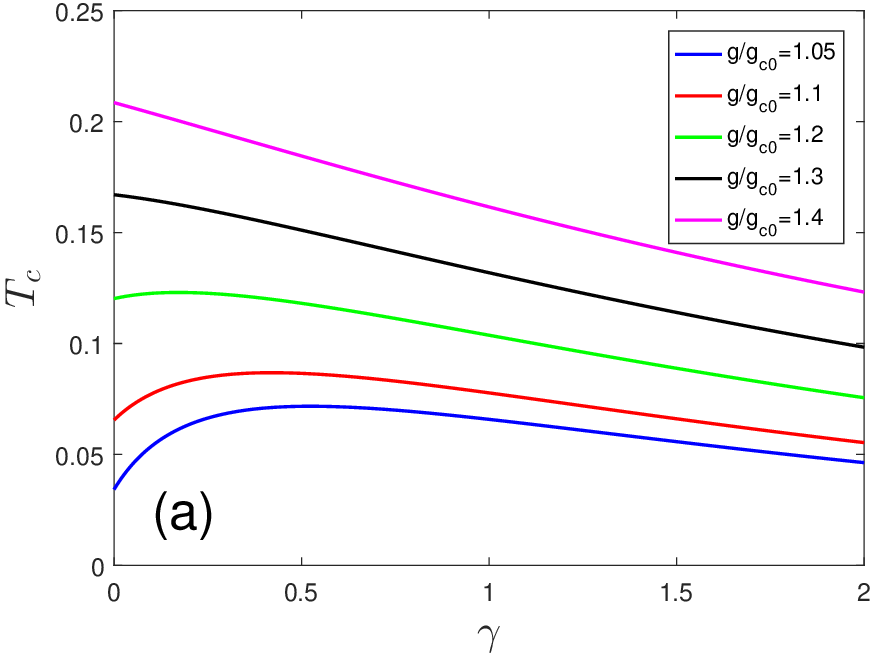}
\includegraphics[width=2.2in]{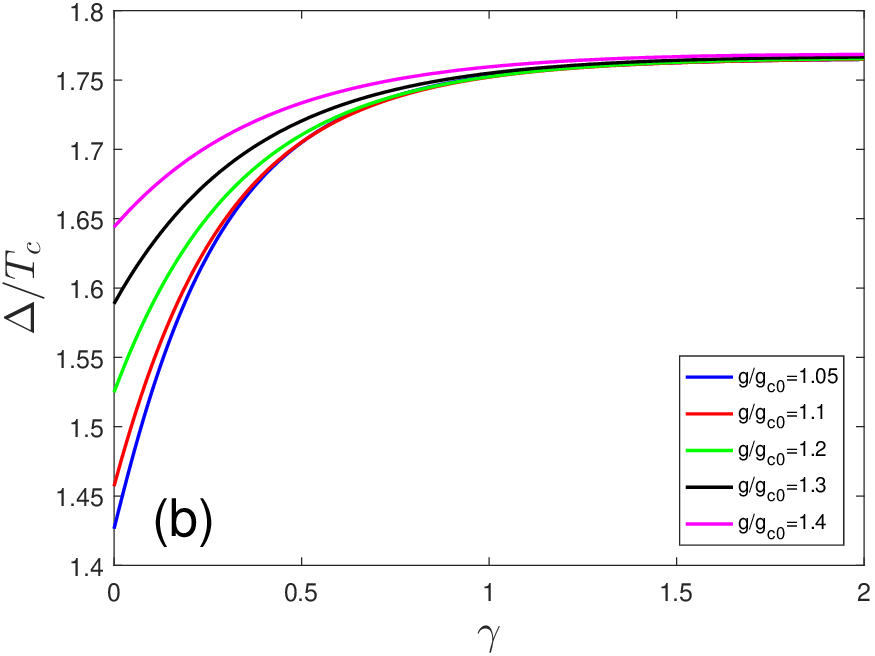}
\caption{(a) Dependence of $T_{c}$ and (b) $\Delta/T_{c}$ on $\gamma$ at different values of $g$ for 2D DSM.
\label{Fig:TcDSM2D}}
\end{figure}

How RSP affects the ratio $\Delta/T_{c}$ is an interesting question. In this
subsection, we show the results obtained by AG method for 2D DSM. The results for
other SMs could be calculated similarly. For 2D DSM, the
self-consistent equations for $T_{c}$ are given by
\begin{eqnarray}
A&=&1+\frac{\gamma}{2}A\ln\left(1+\frac{1}{A^{2}
(2n+1)^{2}\pi^{2} T_{c}^{2}}\right),
\\
1&=&\frac{g}{g_{c0}}T_{c}\sum_{n}A
\ln\left(1+\frac{1}{A^{2}(2n+1)^{2}\pi^{2} T_{c}^{2}}\right).
\end{eqnarray}
The relation between $T_{c}$ and $\gamma$ is displayed in figure~\ref{Fig:TcDSM2D}(a).
Comparing figure~\ref{Fig:TcDSM2D}(a) with figure~\ref{Fig:Gapgamma3D}, we can find that
relation between $T_{c}$ and $\gamma$ has similar characteristic with relation between $\Delta$ on $\gamma$. Dependence of $\Delta/T_{c}$ on $\gamma$ is shown in figure~\ref{Fig:TcDSM2D}(b). According to figure~\ref{Fig:TcDSM2D}(b), $\Delta/T_{c}$ increases with growing of $\gamma$.

\subsection{Anderson localization}

DSMs have obvious different behaviors
under the influence of RSP comparing with traditional metals.
For 2D contentional metals, arbitrarily weak RSP drives
the system to Anderson insulator (AI) \cite{Abrahams79}. However, for 2D DSM with a single Dirac cone, the system is robust against  Anderson localization under RSP \cite{Evers08, Ostrovsky07, Ryu07, Nomura07}.

For 3D  conventional metals, there are only two phases under the influence of RSP, namely CDM and AI \cite{Abrahams79}. However, for 3D DSM, there are three phases
under the influence of RSP, namely SM, CDM, and AI \cite{Pixley15, Liu16}.  We should notice that
the critical value $\gamma_{c}=1$ is corresponding to the QCP from
SM to CDM \cite{Pixley15, Liu16, Goswami11, Roy14}.  The transition from CDM to AI occurs at another critical value $\gamma_{c}^{\mathrm{AI}}$ which is much larger than $\gamma_{c}$ \cite{Pixley15, Liu16}. In a wide range
value of $\gamma$ around $\gamma_{c}$, the system is still in SM or CDM phase, but
not AI phase. Thus, in this case, the wave functions are not localized, but still extended. Then, our calculation is valid for a wide range of disorder strength around
$\gamma_{c}=1$. If the disorder strength is very large, i.e. $\gamma>\gamma_{c}^{\mathrm{AI}}$, Anderson localization appears, and our calculation becomes invalid.

\section{Summary and conclusion\label{Sec:SummaryAndDiscussion}}

In summary, we have studied the influence of random chemical
potential on the fate of $s$-wave superconductivity in 2D DSM by
using the AG diagrammatic approach along with its proper
generalization. It is found that an arbitrarily weak attraction
suffices to trigger Cooper pairing instability. In the case of weak
attraction, the magnitude of SC gap first increases with growing
disorder strength and then decreases once the disorder strength
exceeds a critical value. For relatively strong attraction, the gap
decreases monotonously as disorder gets stronger. To obtain a
quantitatively more reliable result, we have gone beyond the
original AG approximation, and taken into account the vertex
correction to the fermion-disorder coupling. Our finding is that,
including vertex correction does not change the qualitative behavior
of superconductivity in 2D DSM with random chemical potential.
However, for weak pairing interaction and weak disorder, the
disorder-induced enhancement of SC gap size becomes more significant
in the presence of vertex correction. Therefore, the conclusion that
superconductivity in 2D DSM is promoted by random chemical potential
\cite{Nandkishore13} is robust.

We then have applied the AG method and its generalization to
investigate the fate of $s$-wave superconductivity in other
analogous materials, including 3D DSM, 2D semi-DSM, and bilayer
graphene. For 3D DSM, we have found that the critical pairing
interaction strength $g_c$ is reduced to a smaller value by weak
random chemical potential, and thus there is still a QCP separating
the SM and SC phases. This QCP provides an ideal platform to study
the rich quantum critical phenomena. Nevertheless, when random
chemical potential becomes sufficiently strong, the critical value
$g_c$ vanishes, and superconductivity is achieved no matter how weak
the pairing interaction. In both cases, we see that
superconductivity is promoted by random chemical potential.

For 2D semi-DSM, the disorder effect on the $s$-wave
superconductivity is nearly the same as that of 2D DSM. In
particular, superconductivity can be induced by arbitrarily weak
attraction if the system contains random chemical potential. When
the vertex correction is considered, the promotion of
superconductivity by disorder in the weak attraction regime also
becomes more obvious. However, the vertex correction does not modify
the qualitative results obtained by using the original AG
approximation.

Comparing to the above three types of SMs, the bilayer graphene is
spectacular since its zero-energy DOS takes a finite value at the
Fermi level. As the disorder strength increases, the magnitude of
the SC gap decreases, yet at a very low speed. Apparently, such
behavior is in sharp contrast to that of 2D DSM, 3D DSM, and 2D
semi-DSM, where the zero-energy DOS vanishes in the clean limit and
acquires a finite value only when the system contains random
chemical potential.

Recently, Ozfidan \emph{et al.} studied the influence of magnetic
impurity on superconductivity in 2D DSM by using the AG method
\cite{Ozfidan16}, and found a gapless helical SC state.
It would be interesting in the future to investigate whether the
vertex correction makes an important contribution to the physical
effects of magnetic disorder on superconductivity.

\ack{We would acknowledge the financial support by the National Key R\&D Program of
China under Grants 2016YFA0300404 and 2017YFA0403600, and the financial support by the National
Natural Science Foundation of China under Grant Nos. 11574285,
11504379, 11674327, U1532267, and U1832209. J.R.W. is also supported by the
Natural Science Foundation of Anhui Province under Grant No.
1608085MA19.}

\end{document}